\documentclass[prc,twocolumn,showpacs,showkeys,preprintnumbers,superscriptaddress,amsmath,amssymb,nofootinbib]{revtex4}

\usepackage[dvips]{graphicx,psfrag}
\usepackage{dcolumn}
\usepackage{bm}
\usepackage{amsmath, amssymb}

\usepackage{color}
\usepackage{ulem} 

\renewcommand\sout[1]{}

\newcommand{\Psfig}[2]{\includegraphics[width=#1]{#2}}

\def\be{\begin{equation}}
\def\ee{\end{equation}}
\def\bc{\begin{center}}
\def\ec{\end{center}}

\def\LamFOF{\Lambda \text{(1405)}}
\def\SigTEF{\Sigma \text{(1385)}}

\def\LamStar{\Lambda ^{\ast}}
\def\SigStar{\Sigma ^{\ast}}
\def\SigStarz{\Sigma ^{\ast 0}}
\def\SigStarm{\Sigma ^{\ast -}}

\def\KbarN{\bar{K} N}

\def\mev{\text{ MeV}}
\def\gev{\text{ GeV}}
\def\fm{\text{ fm}}

\def\HeT{{}^{3}\text{He}}
\def\HeF{{}^{4}\text{He}}

\begin{document}

\preprint{YITP-12-32, J-PARC-TH-004}%

\title{Branching ratios of mesonic and nonmesonic antikaon absorptions
  in nuclear medium}

\author{Takayasu Sekihara} 
\altaffiliation[The present address is ]{Institute of Particle and
  Nuclear Studies, High Energy Accelerator Research Organization
  (KEK), 1-1, Oho, Ibaraki 305-0801, Japan.}
\affiliation{Department of Physics, Tokyo Institute of Technology,
  Tokyo 152-8551, Japan}

\author{Junko Yamagata-Sekihara}
\altaffiliation[The present address is ]{Institute of Particle and
  Nuclear Studies, High Energy Accelerator Research Organization
  (KEK), 1-1, Oho, Ibaraki 305-0801, Japan.}
\affiliation{Departamento de F\'{\i}sica Te\'orica and IFIC,
  Centro Mixto Universidad de Valencia-CSIC,
  Institutos de Investigaci\'on de Paterna, Aptdo. 22085, 46071 
  Valencia, Spain}

\author{Daisuke Jido} 
\affiliation{Yukawa Institute for Theoretical Physics, 
  Kyoto University, Kyoto 606-8502, Japan}
\affiliation{J-PARC Branch, KEK Theory Center,
  Institute of Particle and Nuclear Studies, 
  High Energy Accelerator Research Organization (KEK),
  203-1, Shirakata, Tokai, Ibaraki, 319-1106, Japan}

\author{Yoshiko Kanada-En'yo} 
\affiliation{Department of Physics, Kyoto University, Kyoto 606-8502, Japan}

\date{\today}

\begin{abstract}
  The branching ratios of $K^{-}$ absorption in nuclear matter are
  theoretically investigated in order to understand the mechanism of
  $K^{-}$ absorption into nuclei.  For this purpose mesonic and
  nonmesonic absorption potentials are evaluated as functions of
  nuclear density, the kaon momentum and energy from one- and two-body
  $K^{-}$ self-energy, respectively.  By using a chiral unitary
  approach for the $s$-wave $\KbarN$ amplitude we find that both the
  mesonic and nonmesonic absorption potentials are dominated by the
  $\LamFOF$ contributions.  The fraction of the mesonic and nonmesonic
  absorptions are evaluated to be respectively about $70 \%$ and $30
  \%$ at the saturation density almost independently on the kaon
  momentum. We also observe different behavior of the branching ratios
  to $\pi ^{+} \Sigma ^{-}$ and $\pi ^{-} \Sigma ^{+}$ channels in
  mesonic absorption due to the interference between $\LamFOF$ and the
  $I=1$ nonresonant background, which is consistent with experimental
  results.  The nonmesonic absorption ratios $[\Lambda p]/ [\Sigma
  ^{0} p]$ and $[\Lambda n]/ [\Sigma ^{0} n]$ are about unity while
  $[\Sigma ^{+}n]/[\Sigma ^{0}p]$ and $[\Sigma ^{-}p]/[\Sigma ^{0}n]$
  are about two due to the $\LamFOF$ dominance in absorption.  Taking
  into account the kaon momenta and energies, the absorption
  potentials become weaker due to the downward shift of the initial
  $K^{-}N$ two-body energy, but this does not drastirally change the
  nonmesonic fraction.  The $\SigTEF$ contribution in the $p$-wave
  $\KbarN$ amplitude is examined and found to be very small compared
  to the $\LamFOF$ contribution in slow $K^{-}$ absorption.
\end{abstract}


\maketitle

\section{Introduction}
\label{sec:introduction}

Interaction between antikaon ($\bar{K}$) and nucleon ($N$) is one of
the most important clues for strangeness nuclear physics.  The
$\KbarN$ interaction in the $I=0$ channel is strongly attractive at
low energies and dynamically generates $\LamFOF$ as a quasi-bound
state of $\KbarN$, which couples to $\pi \Sigma$ as a decay
mode~\cite{Dalitz:1967fp} (see also~\cite{Hyodo:2011ur}).  The
attractive interaction between $\KbarN$ stimulates recent theoretical
studies on $\bar{K}$ few-nucleon systems bound mainly by the
strong interaction (kaonic nuclei)~\cite{Kishimoto:1999yj,
  Akaishi:2002bg, Yamagata:2006sm,Shevchenko:2006xy,Ikeda:2007nz,
  Dote:2008in,Wycech:2008wf,Bayar:2011qj} and further nuclear systems
with kaons such as $\bar{K} K N$~\cite{Jido:2008kp,
  MartinezTorres:2008kh,MartinezTorres:2010zv,Xie:2010ig} and $\bar{K}
\bar{K} N$~\cite{KanadaEn'yo:2008wm}.  The $\KbarN$ interaction is
also related to the in-medium properties of
$\bar{K}$~\cite{Koch:1994mj,
  Waas:1996xh,Waas:1996fy,Waas:1997pe,Lutz:1997wt,Ramos:1999ku}, which
is a key to the kaon behavior in high dense
matter~\cite{Kaplan:1986yq}.  However, at present the
low-energy $\KbarN$ interaction is not well-understood especially in
its subthreshold regions.

An important tool to study the phenomenological $\bar{K}$-nucleus
interaction at low energies including the $\KbarN$ interaction is
kaonic atoms, which are coulombic bound states of $K^-$-nucleus with
influence of strong interaction.  Kaonic atoms have attracted much
attention both experimentally and theoretically, because they provide
unique information on strong interaction between nucleus and $K^{-}$
at zero momentum from their binding energies and decay
widths~\cite{Batty:1997zp,Friedman:2007zza}.  In earlier works, the
branching ratios of $K^{-}$-nucleus absorption at rest had been
experimentally investigated from the 1960's to the 1970's by using
emulsions and bubble chambers with, for example,
hydrogen~\cite{Tovee:1971ga, Nowak:1978au},
deuterium~\cite{Veirs:1970fs}, $\HeF$~\cite{Katz:1970ng}, and heavier
nucleus~\cite{VanderVeldeWilquet:1977rw}.  As a result, it was found
that probability to observe nonpionic final state is as large as $20
\%$ per stopped $K^{-}$ for $\HeF$ and heavier
nuclei~\cite{VanderVeldeWilquet:1977rw} while it is $\sim 1 \%$ for
deuterium~\cite{Veirs:1970fs}.  The fraction of the nonpionic final
state for kaon absorption by $\HeF$ was theoretically studied in
Ref.~\cite{Onaga:1989dn}.  A detailed analysis of the branching ratios
for stopped $K^{-}$ on $\HeF$ was performed in Ref.~\cite{Katz:1970ng}
and the authors reported, for example, the absorption ratio
$R_{+-}\equiv [\pi ^{-} \Sigma ^{+}] / [\pi ^{+} \Sigma ^{-}] = 1.8
\pm 0.5$, which is larger than smaller systems, such as $\approx 0.42$
for hydrogen~\cite{Tovee:1971ga,Nowak:1978au} and $\approx 0.85$ for
deuterium~\cite{Katz:1970ng}.  The ratio $R_{+-}$ is also studied in
Ref.~\cite{Agnello:2011iq} for $p$-shell nuclei and $R_{+-} =
1.2$--$1.5$ is obtained. It was theoretically suggested in
Refs.~\cite{Staronski:1987he,Ohnishi:1997pt} that the ratio $R_{+-}$
strongly reflects the in-medium properties of $\LamFOF$.  In recent
works, the energy shift and width of the $1s$ state in kaonic hydrogen
is experimentally extracted in Refs.~\cite{Iwasaki:1997wf,Ito:1998yi,
  Beer:2005qi,Bazzi:2011zj,Bazzi:2012eq}, which are followed by
theoretical improvements of the $\KbarN$ interaction around and below
the threshold in Refs.~\cite{Ikeda:2011pi,Ikeda:2012au, Mai:2012dt}.
In Refs.~\cite{Cieply:2011yz, Cieply:2011fy, Friedman:2012pc,
  Gazda:2012zz} theoretical analyses of kaonic atoms data including
heavy nuclei are performed with subthreshold in-medium $\KbarN$
scattering amplitudes and $K^{-}$-nucleus potentials by strong
interaction as well as propeties of kaonic nuclei are discussed.
Searches for kaonic nuclei~\cite{Agnello:2005qj,Agnello:2007aa,
  Suzuki:2007kn} were done in stopped $K^{-}$ experiments by detecting
$\Lambda$-nucleus correlations in the final state of stopped $K^{-}$
absorption reactions, motivated by the deeply bound kaonic nuclei
predicted in Ref.~\cite{Akaishi:2002bg}. However there is no clear
evidence yet and further there are discussions on alternative
explanations for the peaks observed in experiments~\cite{Oset:2005sn,
  Magas:2006fn, Magas:2008bp}.

One considerable feature of the $K^{-}$-nucleus absorption process at
rest is that the energy of the $K^{-} N$ two-body system in the
initial state can go below the threshold due to the off-shellness of
the bound nucleon inside the nucleus.  This leads to the expectation
that the absorption pattern is closely related to the $\KbarN$
dynamics below the threshold.  Especially there are two hyperon
resonances, $\LamFOF$ ($\LamStar$) and $\SigTEF$ ($\SigStar$) below
the $\KbarN$ threshold, hence it is natural to consider that their
contributions to the absorption process are important.  Since
$\LamStar$ strongly couples to the $\KbarN$ channel in $s$ wave,
$\LamStar$ will play the most important role.  Therefore, it is
interesting to construct $\bar{K}$-nucleus interactions from the
$\KbarN$ interaction including $\LamStar$ and $\SigStar$ and to
investigate systematically the branching ratios of the $\bar{K}$-nucleus
systems from viewpoint of the low-energy $\KbarN$ interaction in order
to understand the mechanism of $K^{-}$ absorption in experiments.

Motivated by these observations, in this study we theoretically
investigate the decay pattern of $K^{-}$ in nuclear matter as a
simplified condition for the kaonic atoms by calculating the imaginary
part of the $K^{-}$ self-energy with $\KbarN$ interaction as an input.
We employ chiral dynamics within a unitary framework (chiral unitary
approach)~\cite{Kaiser:1995eg,Oset:1997it, Oller:2000fj, Lutz:2001yb,
  Oset:2001cn,Jido:2003cb} for the $\KbarN$ interaction.  Here we
investigate mesonic and nonmesonic decay by taking into account the
most probable contributions, that is, the one- and two-nucleon
absorption for the mesonic and nonmesonic decay, respectively.
Multi-nucleon interactions for the mesonic decay as well as the more
than three-nucleon interactions for the nonmesonic decay will be
suppressed when the nuclear density is not so high.  In this study we
consider the $K^{-}$ self-energy as a function of the kaon energy and
momentum as well as the nuclear density.  For bound kaons these energy
and momentum are determined self-consistently by the equation of
motion with the energy dependent potential, and the energy shift and
the momentum distribution should be taken into account especially for
deeply bound kaon states as suggested in Refs.~\cite{Cieply:2011yz,
  Cieply:2011fy}. For simplicity, we assume the isospin symmetry and
consider the symmetric nuclear matter, $\rho _{N}= \rho
_{\text{proton}}+ \rho _{\text{neutron}}$ with $\rho _{\text{proton}}=
\rho _{\text{neutron}}$.  The extension to the case of the asymmetric
matter is straightforward.

\begin{figure}[!Ht]
  \centering
  \begin{tabular*}{8.6cm}{@{\extracolsep{\fill}}ccc}
  \includegraphics[scale=0.17]{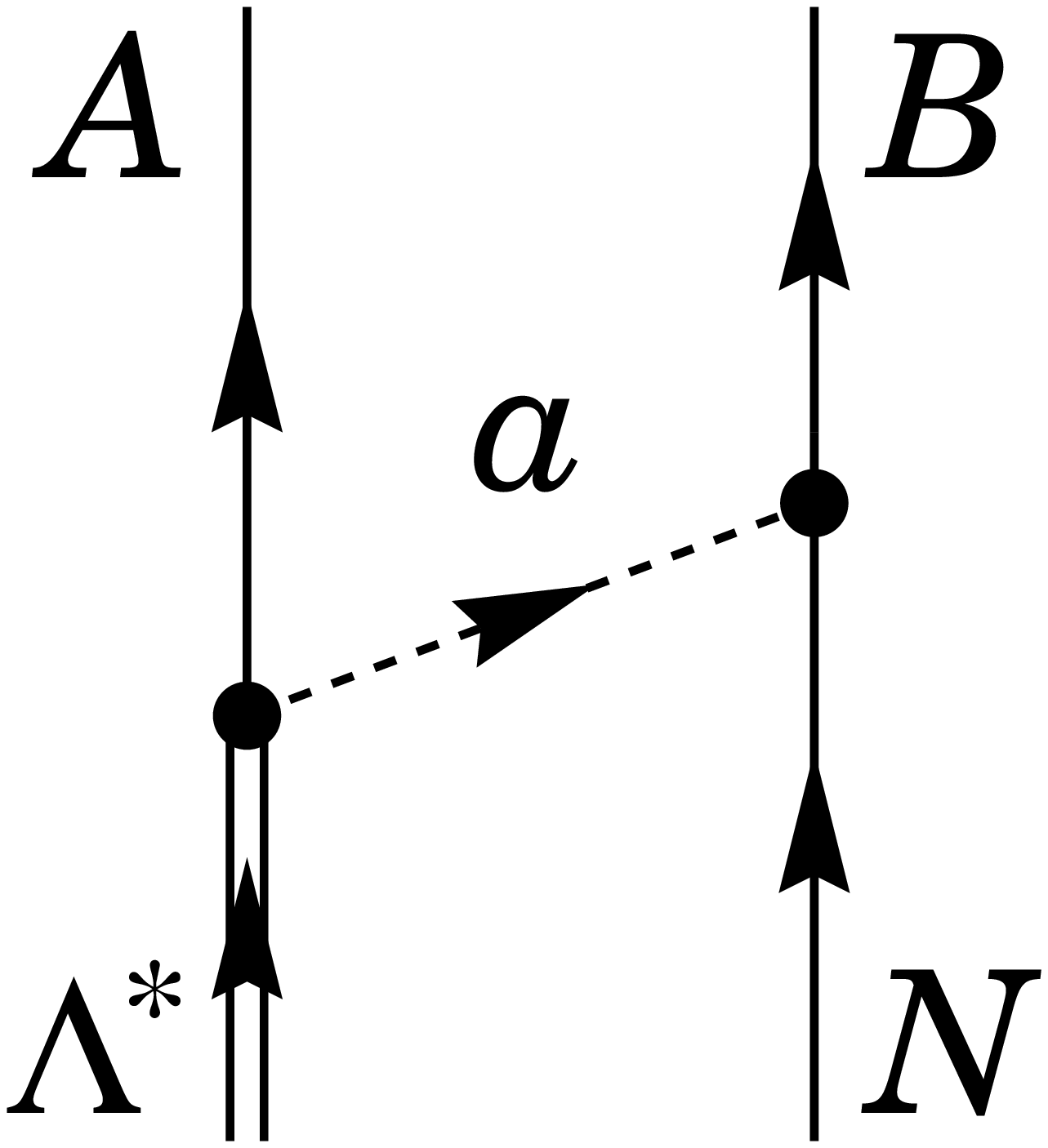} &
  \includegraphics[scale=0.17]{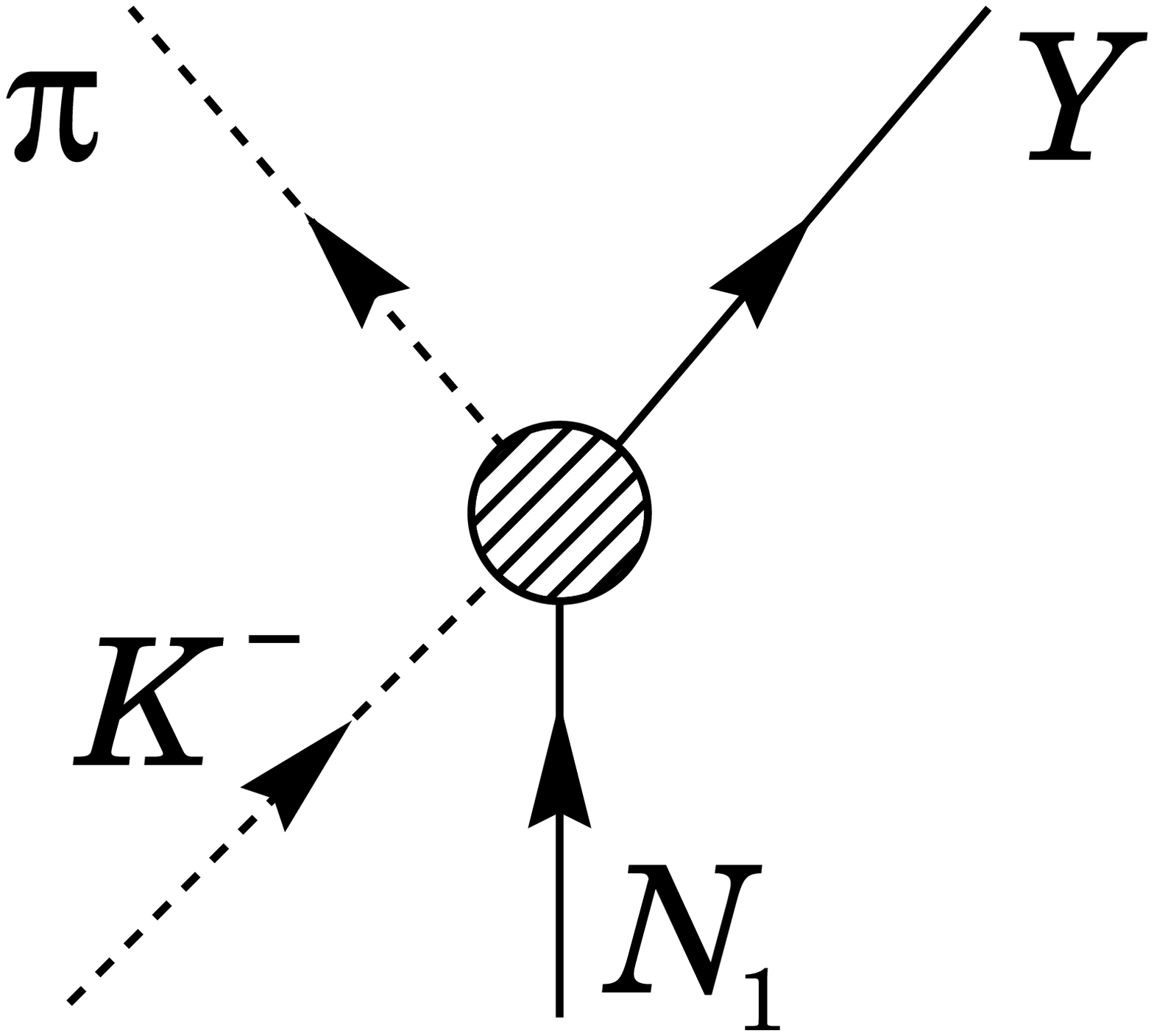} & 
  \includegraphics[scale=0.17]{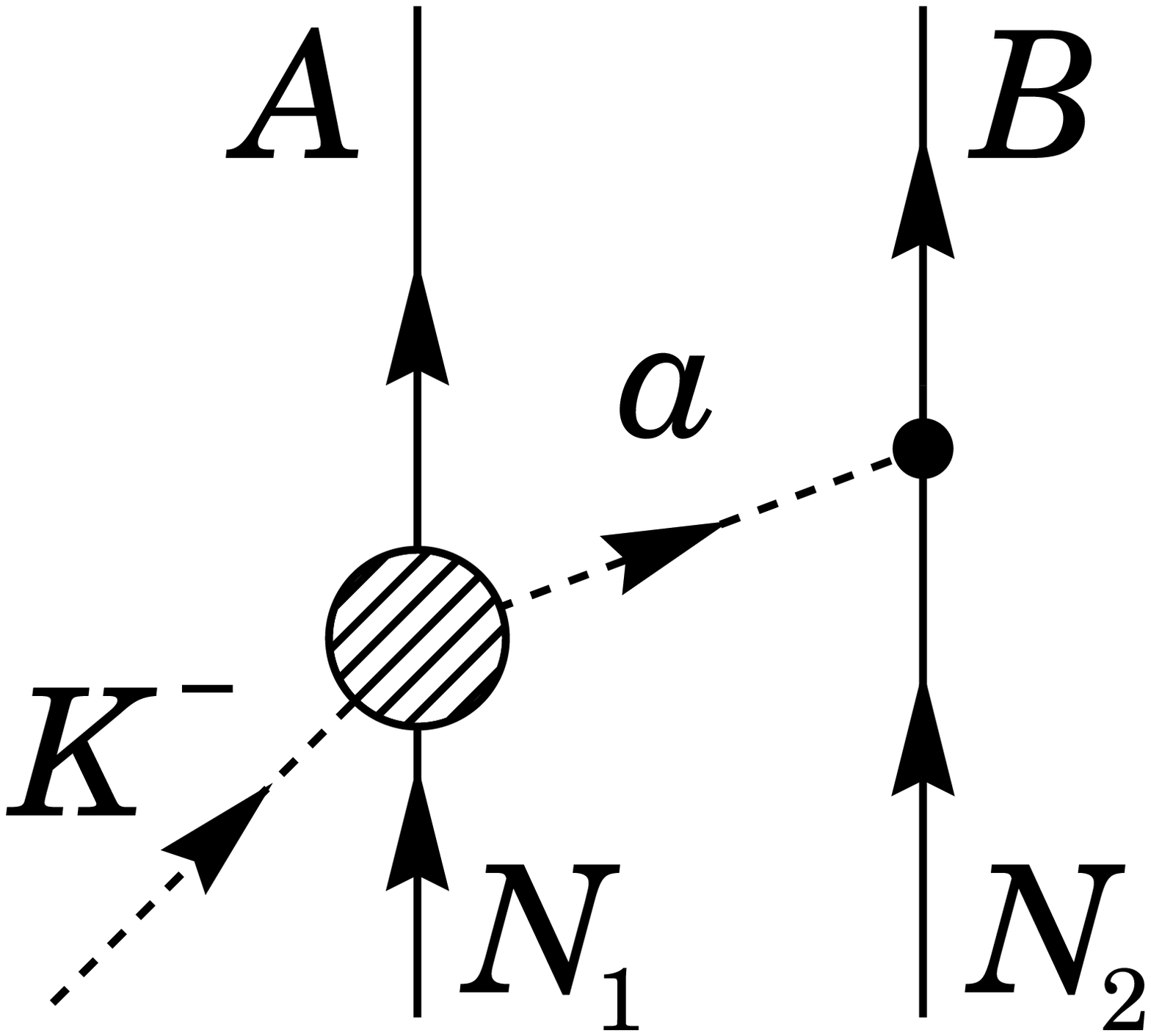} \\
  (a) & (b) & (c)
  \end{tabular*}
  \caption{Feynman diagrams for the decay of $\bar{K}$-nucleus
    systems.  (a) $\LamStar$-induced reaction.  (b) Mesonic absorption
    with an explicit $K^{-}$ in the initial state.  (c) Nonmesonic
    absorption with an explicit $K^{-}$ in the initial state.  In
    diagrams, $A$ and $B$ denote the baryons in the final state and
    $a$ denotes the exchange meson. }
  \label{fig:Feynman-prev}
\end{figure}

The present study is a continuation of the study done in
Ref.~\cite{Sekihara:2009yk}.  In Ref.~\cite{Sekihara:2009yk} we have
discussed nonmesonic decay of $\LamStar$ in a nuclear medium by
employing one-meson exchange model as diagrammatically shown in
Fig.~\ref{fig:Feynman-prev}(a), and we have found that the nonmesonic
decay ratio $\Gamma _{\Lambda N} / \Gamma _{\Sigma ^{0} N}$ strongly
depends on the $\LamStar$ coupling ratio $g_{\KbarN}/g_{\pi \Sigma}$;
especially large $g_{\KbarN}$ coupling leads to the enhancement of
$\Gamma _{\Lambda N}$.  Futhermore, by using the chiral unitary
approach we have found that $\Gamma _{\Lambda N} / \Gamma _{\Sigma
  ^{0} N} \approx 1.2$ almost independently of the nuclear density.
In the previous study it has been assumed that one $\LamStar$ is
created in nuclear matter and the nonmesonic decay pattern of
$\bar{K}$-nucleus bound systems has been discussed in an idealized
condition.  In the present study we consider an explicit $K^{-}$ in
the initial state rather than $\LamStar$ as shown in
Figs.~\ref{fig:Feynman-prev}(b) and (c), which enables us to
investigate the decay of $\bar{K}$-nucleus bound systems in more
realistic conditions.  We discuss how much partition we observe the
$\LamStar$ dominance in $K^{-}$ absorption, which is assumed to be
perfect in our previous study.  In this work we neglect in-medium
effects on mesons, baryons, and hyperon resonances.  It is known that
the Pauli blocking effect on the nucleons makes $\LamStar$ energy
shift above the $\KbarN$ threshold as discussed in
Refs.~\cite{Koch:1994mj,Waas:1996xh,Waas:1996fy, Waas:1997pe}.
Nevertheless, taking into account the in-medium effects on
$\bar{K}$~\cite{Lutz:1997wt} and both on $\bar{K}$ and
$\pi$~\cite{Ramos:1999ku} as dressed propagators, it was suggested
that the in-medium attraction felt by $\bar{K}$ lowers the $\KbarN$
threshold and thus $\LamStar$ position moves to the energy close to
its free-space value.  Although the clear peak of $\LamStar$ would be
dissolved in nuclear matter, we use $\LamStar$ without in-medium
effects as a zeroth order approximation.

This paper is organized as follows. In Sec.~\ref{sec:formulation} we
explain our formulation for the calculation of the mesonic and
nonmesonic absorption potentials for $K^{-}$ in nuclear matter.  We
show our results of the absorption potential with $s$-wave $\KbarN \to
MB$ scattering amplitude including the $\LamStar$ contributions in
Sec.~\ref{sec:3}.  The $\SigStar$ contributions is included in
Sec.~\ref{sec:Sigma1385}.  Section~\ref{sec:summary} is devoted to
summary of this paper.

\section{Formulation}
\label{sec:formulation}

\begin{figure}[!Ht]
  \bc
  \includegraphics[scale=0.17]{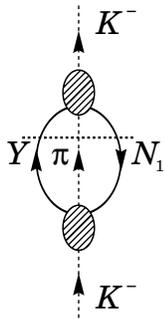}
  \caption{Feynman diagram for the mesonic $K^{-}$ absorption
    processes in nuclear matter.  The possible combination of the
    nucleon ($N_{1}$), hyperon ($Y$), and pion ($\pi$) in the
    intermediate state is given in Table~\ref{tab:Feynman-a}.  The
    shaded ellipses represent the $K^{-}N \to \pi Y$ amplitudes. }
  \label{fig:Feynman-K1}
  \ec
\end{figure}

\begin{figure}[!Ht]
  \bc
  \begin{tabular*}{7.0cm}{@{\extracolsep{\fill}}cc}
    \includegraphics[scale=0.17]{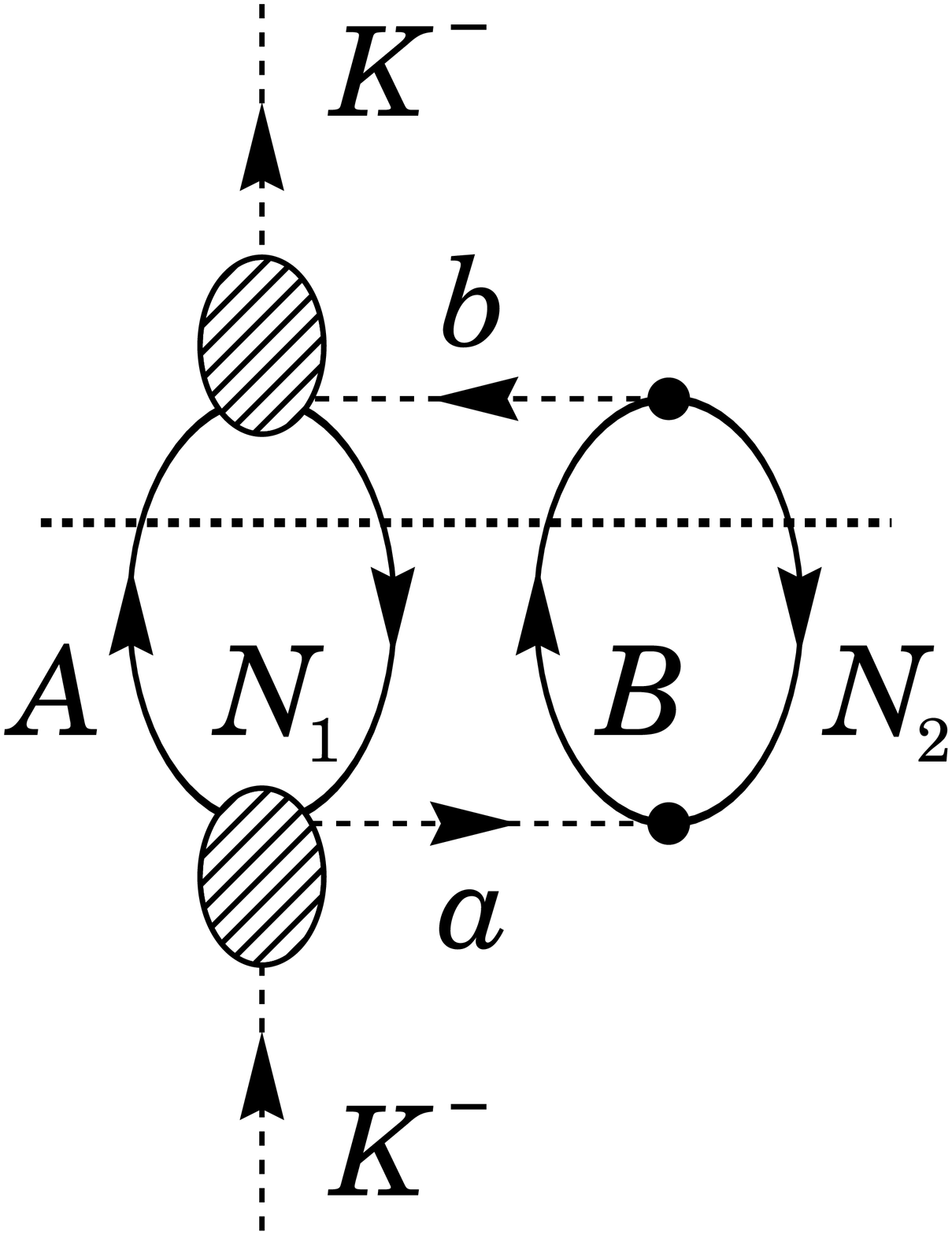}
    &
    \includegraphics[scale=0.17]{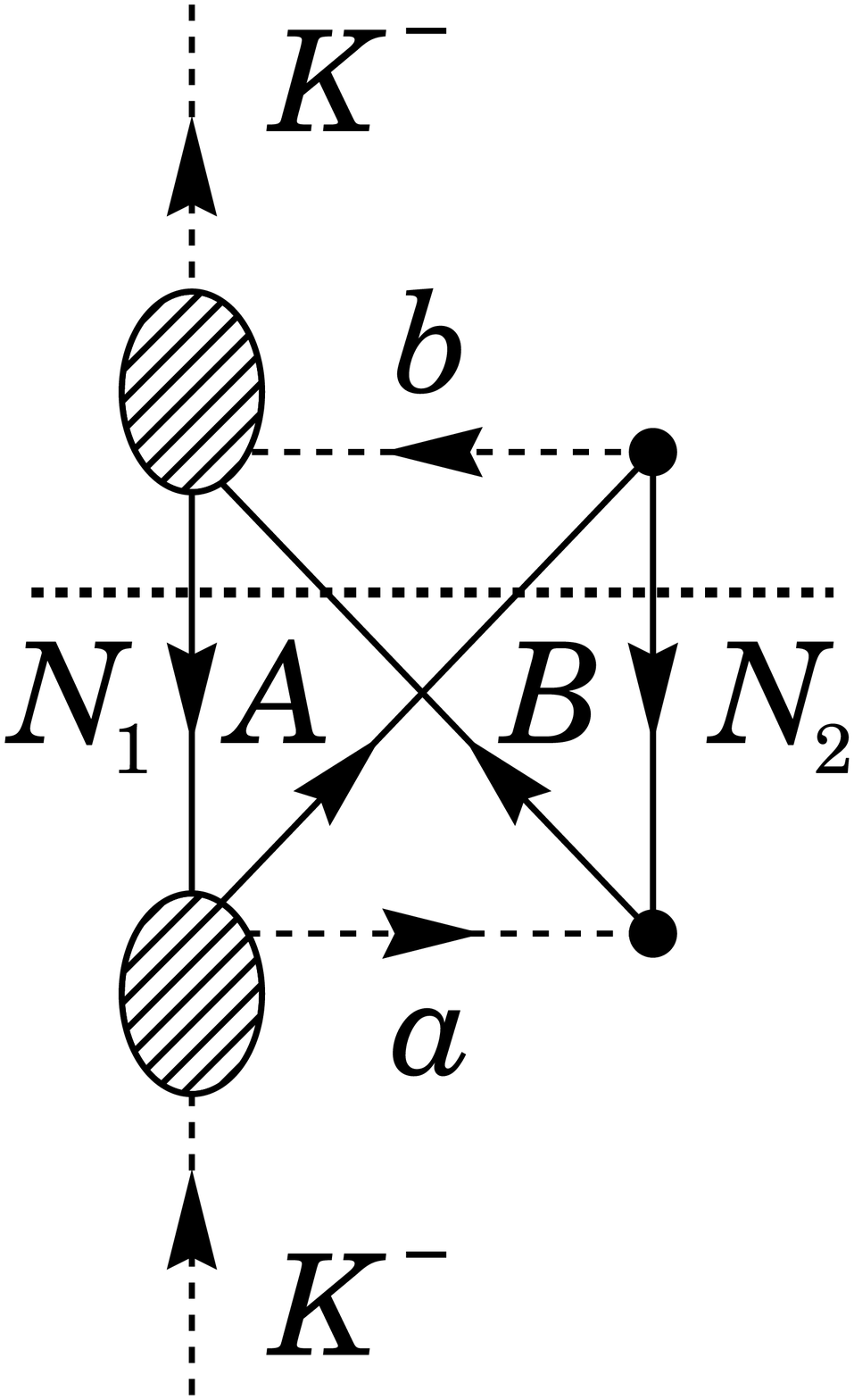}
    \\
    (a) & (b) \vspace{10pt} \\
    \includegraphics[scale=0.17]{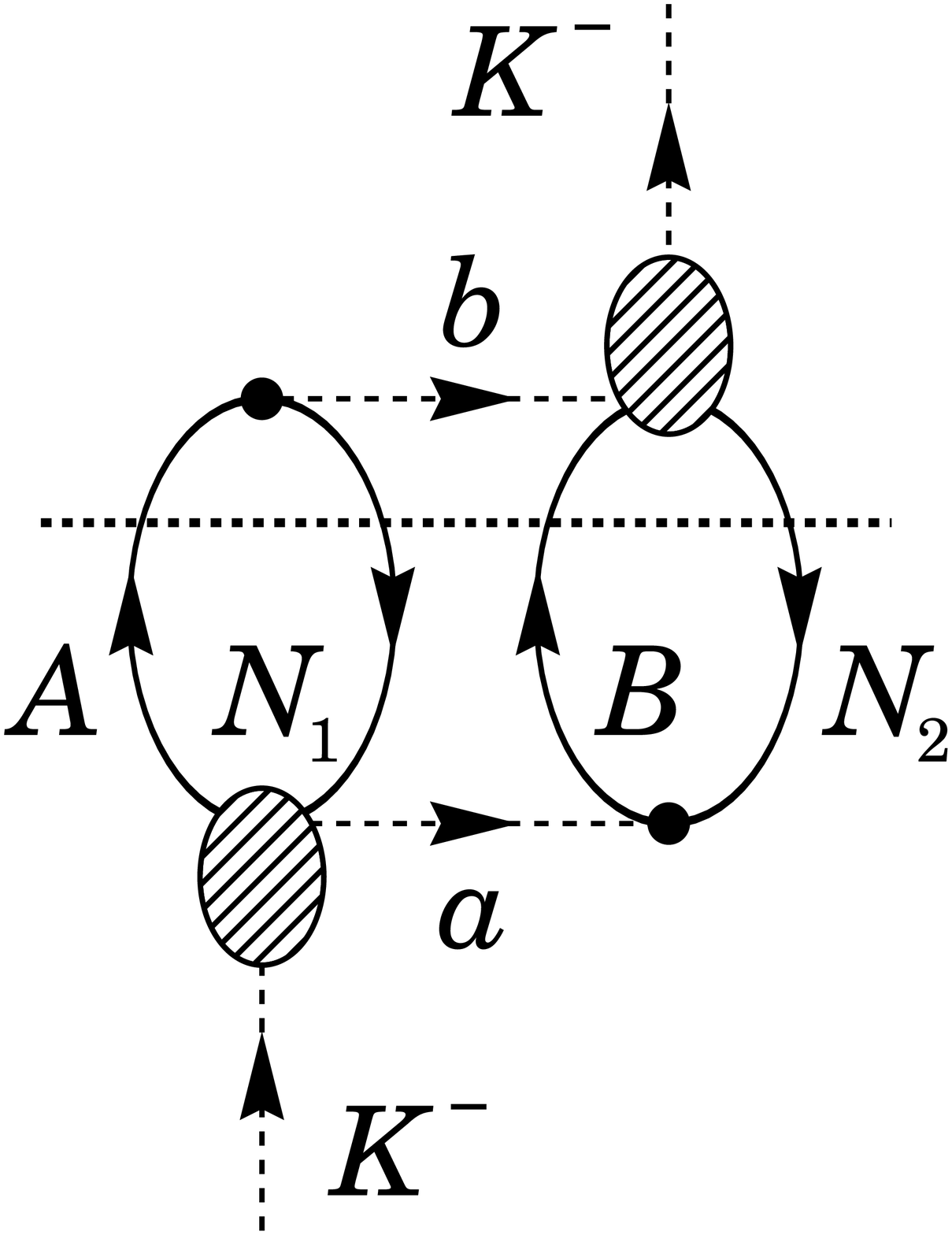}
    & 
    \includegraphics[scale=0.17]{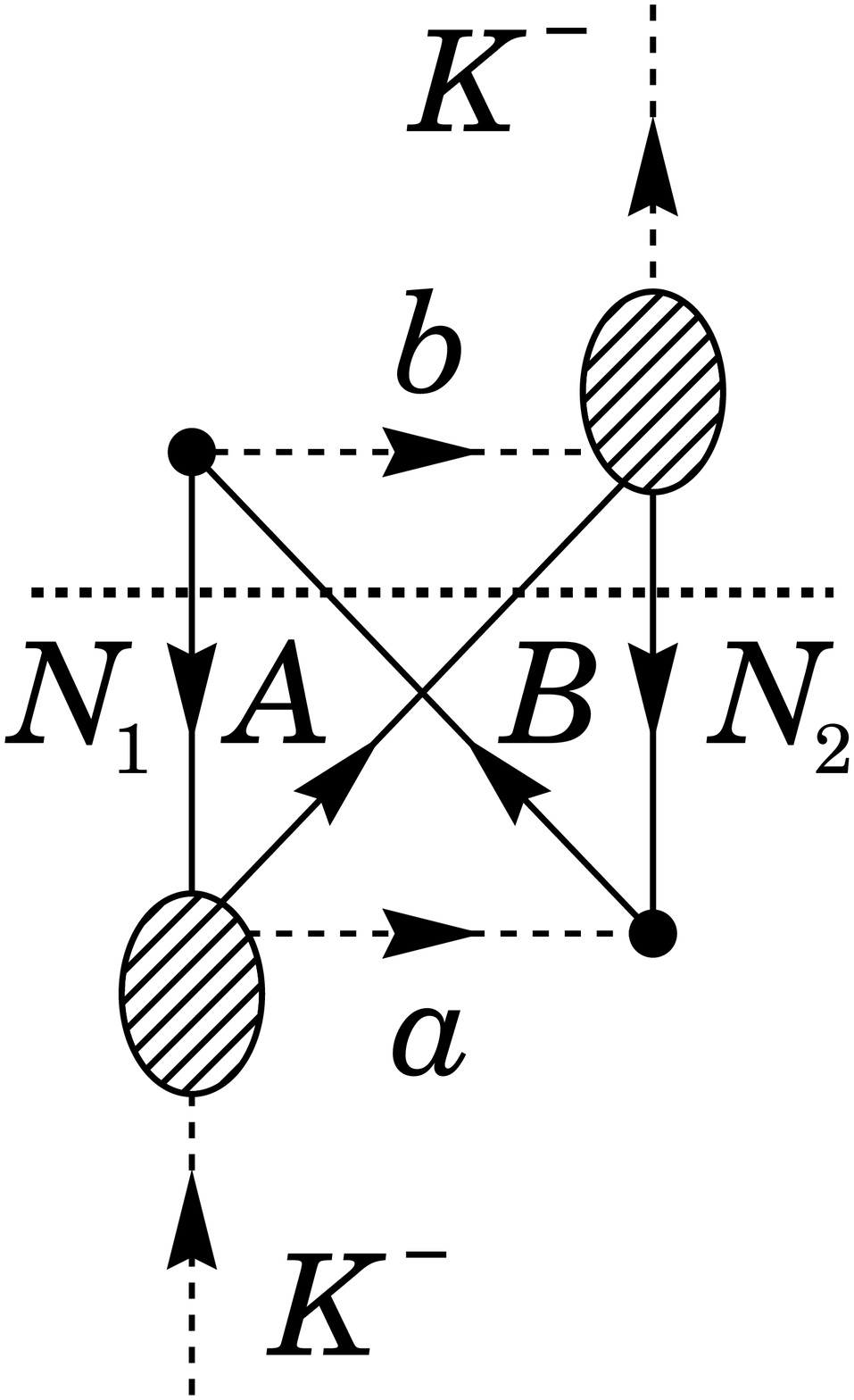}
    \\
    (c) & (d) 
  \end{tabular*}
  \caption{Feynman diagrams for the nonmesonic $K^{-}$ absorption
    processes in nuclear matter.  In diagrams, $N_{1}$ and $N_{2}$
    denote nucleons, $A$ and $B$ baryons, and $a$ and $b$ mesons.  The
    shaded ellipses represent the $K^{-}N \to MB$ amplitudes. }
  \label{fig:Feynman-K2}
  \ec
\end{figure}

In this section we formulate the absorption potential of $K^{-}$ in
uniform nuclear matter which is given by the imaginary part of the
$K^{-}$ self-energy in the medium as a function of nuclear density
$\rho _{N}$ as well as the kaon energy $E_{K^{-}}$ and momentum $p_{K^{-}}$.  
In general, the
potential for $K^{-}$, $V$, can be obtained by evaluating the $K^{-}$
self-energy $U_{K^{-}}$ as,
\begin{equation} 
2 m_{\bar{K}} V = U_{K^{-}} ,
\label{eq:2mV=U}
\end{equation}
with the antikaon mass $m_{\bar{K}}$.  The imaginary part of the
potential $V$ represents the contribution from $K^{-}$ absorption,
\begin{equation}
\text{Im} V = 
\frac{1}{2 m_{\bar{K}}} \text{Im} U_{K^{-}} , 
\end{equation}
in which we are interested here.  In evaluation of the imaginary part
of the self-energy $U_{K^{-}}$ we use the Cutkosky rule.

In this study we discuss the mesonic and nonmesonic absorption
processes for $K^{-}$ in nuclear matter as one- and two-body
absorption by considering diagrams shown in
Figs.~\ref{fig:Feynman-prev}(b) and (c), respectively.  These are the
most kinematically probable contributions to $K^{-}$ absorption.  The
mesonic absorption potential is evaluated from the self-energy of the
Feynman diagram in Fig.~\ref{fig:Feynman-K1}.  For the nonmesonic
absorption, on the other hand, we take one-meson exchange model where
the Nambu-Goldstone bosons are exchanged between the baryons, as
diagrammatically shown in Fig.~\ref{fig:Feynman-K2}.  The sum of the
two contributions gives the total $K^{-}$ self-energy,
\begin{equation}
U_{K^{-}} = U_{K^{-}}^{\text{one}} + U_{K^{-}}^{\text{two}} , 
\end{equation}
where $U_{K^{-}}^{\text{one}}$ and $U_{K^{-}}^{\text{two}}$ represent
the one- and two-body self-energy, respectively.

The $K^{-}$ absorption potential is calculated as a function of the
nuclear density $\rho _{N}$.  In our study we describe nuclear
matter by using the Thomas-Fermi approximation.  In this approximation
a bound nucleon with momentum $p$ has energy
\begin{equation}
E_{N} = M_{N} + \frac{p^{2}}{2M_{N}} + v_{N} , 
\quad 
v_{N} \equiv -
\frac{k_{\text{F}}^{2}}{2 M_{N}} , 
\label{eq:ThomasFermi}
\end{equation}
where $M_{N}$ is the nucleon mass and $v_{N}$ the potential energy for
the nucleon with the Fermi momentum $k_{\text{F}}=(3 \pi ^{2} \rho
_{N} /2)^{1/3}$.  The nucleon momentum $p$ can take a value from $0$
to $k_{\text{F}}$.  Since we consider symmetric nuclear matter, we
have $k_{\text{F}}(\text{proton})=k_{\text{F}}(\text{neutron})$.  The
potential is also a function of the kaon energy and momentum, which
are external variables of the self-energy.  In contrast, if one
considers bound states of a kaon, one has to calculate the
Schr\"{o}dinger or Klein-Gordon equation with this
energy-momentum-dependent potential self-consistently.  Thus, the
potential for the bound kaon should be evaluated with the energy of
the bound kaon and the momentum distribution of the bound state.  One
of the ways to implement the energy and momentum dependence into the
potential for the bound state was suggested in
Refs.~\cite{Cieply:2011yz, Cieply:2011fy}.

One important feature of $K^{-}$ absorption in nuclei is that the
center-of-mass energy of the $K^{-}N$ pair in nuclear matter can go
below the threshold value, $m_{\bar{K}}+M_{N}$, due to the
off-shellness of the bound nucleon. This can be easily seen by
evaluating the center-of-mass energy from the free $K^{-}$ at rest and
bound $N$ momenta, $p_{K^{-}}^{\mu}=(m_{\bar{K}}, \, \bm{0})$ and
$p_{N_{\text{in}}}^{\mu}=(E_{N}, \, \bm{p})$;
$W=\sqrt{(E_{N}+m_{\bar{K}})^2 - \bm{p}^2} < m_{\bar{K}} + M_{N}$
because of $E_{N} \le M_{N}$ for the bound nucleon.  For a kaon with a
finite momentum and a binding energy, the two-body energy $W$ shifts
farther downward due to the off-shellness of the kaon.  We also note
that, since the momentum of the bound nucleon take a value from $0$ up
to $k_{\rm F}$, the span of the $K^{-}$-$N$ pair energy depends on the
density.  At certain density around the saturation density, the
$K^{-}$-$N$ energy goes down around the hyperon resonances ($\LamStar$
and $\SigStar$) sitting below the $\KbarN$ threshold.  Thus, for these
densities, they are expected to give important contributions to the
$K^{-} N \to MB$ transition amplitudes, which are represented as
shaded ellipses in Figs.~\ref{fig:Feynman-K1} and
\ref{fig:Feynman-K2}, and play a crucial role for the absorption
pattern.

In order to describe the $s$-wave $K^{-}N \to MB$ transition
amplitudes around the $\KbarN$ threshold, we use the so-called chiral
unitary approach~\cite{Kaiser:1995eg,Oset:1997it, Oller:2000fj,
  Oset:2001cn,Jido:2003cb}, which is based on chiral dynamics within a
unitary framework.  Using the parameter set in
Ref.~\cite{Oset:2001cn}, which is fixed by the branching ratios of
$K^{-}p$ at threshold observed with the kaonic
hydrogen~\cite{Tovee:1971ga,Nowak:1978au}, we can reproduce well the
low-energy $\KbarN$ scatterings in $s$ wave and dynamically generate
the $\LamStar$ resonance.  In the chiral unitary approach the
$\LamStar$ peak position initiated from the $\KbarN$ channel is
evaluated to be about $1420 \mev$ instead of the nominal $1405
\mev$~\cite{Jido:2003cb}, which is consistent with the experimental
observation~\cite{Braun:1977wd,Jido:2009jf}.  The details of the
formulation of the chiral unitary approach used here are given in
Ref.~\cite{Sekihara:2010uz}.  The chiral unitary approach is suitable
for our study of $K^{-}$ absorption since this approach automatically
includes the nonresonant background contributions as well as the
$\LamStar$ contribution in the scattering amplitude.  Here we do not
take into account the in-medium effects on the amplitudes determined
by the chiral unitary approach.  We will also examine the $\SigStar$
contribution in $\KbarN$ $p$-wave in Sec.~\ref{sec:Sigma1385} by
introducing a simple Breit-Wigner scattering amplitude for $\SigStar$.

\begin{table}[t]
  \caption{\label{tab:Feynman-a} 
    Channels of the intermediate states in Fig.~\ref{fig:Feynman-K1}. }
  \begin{ruledtabular}
    \begin{tabular}{ccc} 
      $N_{1}$ & $\pi$ & $Y$ \\ \hline
      $p$ & \rule[0pt]{0pt}{9pt} $\pi ^{0}$ & $\Sigma ^{0}$ \\
      & $\pi ^{+}$ & $\Sigma ^{-}$ \\
      & $\pi ^{-}$ & $\Sigma ^{+}$ \\
      & $\pi ^{0}$ & $\Lambda$ \\
      \hline 
      $n$ & \rule[0pt]{0pt}{9pt} $\pi ^{-}$ & $\Lambda$ \\
      & $\pi ^{-}$ & $\Sigma ^{0}$ \\
      & $\pi ^{0}$ & $\Sigma ^{-}$ \\
    \end{tabular}
  \end{ruledtabular}
\end{table}%

Now let us formulate the $K^{-}$ potential for the mesonic absorption,
which is calculated by considering $K^{-}N_{1} \to \pi Y$ process for
the in-medium nucleon $N_{1}$ diagrammatically shown in
Fig.~\ref{fig:Feynman-K1}.  In this study we use a symbol $\mu _{1} =
(\bm{p}_{1}, \, \chi _{1})$ to denote collectively the initial-state
nucleon momentum $\bm{p}_{1}$ and its spin $\chi _{1}$, and we assume
the isospin symmetry.  The cut amplitude for the mesonic process is
given as,
\begin{align}
2 \text{Im} U_{K^{-}}^{\text{one}} 
& = 
- 
\int ^{k_{\text{F}}} \frac{d^{3} p_{1}}{(2 \pi)^{3}} g_{N} 
\overline{\sum_{\lambda}} \sum _{\lambda ^{\prime}} 
\sum _{(\pi , Y)}
\gamma _{\pi Y} (\mu _{1} ; \, k_{\text{F}}) , 
\label{eq:U-one}
\end{align}
with the reaction rate for the $K^{-}N_{1}\to \pi Y$ process, 
\begin{align}
\gamma _{\pi Y} (\mu _{1} ; \, k_{\text{F}}) 
\equiv 
& \int d \Phi _{\pi Y} 
\left | 
\chi _{Y}^{\dagger} 
{\cal T}_{\pi Y} 
\chi _{1} 
\right | ^{2} 
\nonumber \\
& \times (2 \pi)^{4} \delta ^{4} (p_{K^{-}} + p_{1} - p_{\pi} - p_{Y}) . 
\label{eq:gamma-one}
\end{align}
Here $g_{N}=2$ is the degenerate number of the nucleon for each
momentum in nuclear matter (spin up and down), the phase space of the
intermediate on-shell state ($\pi Y$) $d \Phi _{\pi Y}$,
\begin{equation}
d \Phi _{\pi Y} \equiv 
\frac{d^{3} p_{\pi}}{(2 \pi)^{3}} \frac{1}{2 \omega _{\pi}} 
\frac{d^{3} p_{Y}}{(2 \pi)^{3}} \frac{2 M_{Y}}{2 E_{Y}}, 
\end{equation}
the $K^{-}N_{1}\to \pi Y$ scattering amplitude ${\cal T}_{\pi Y}$,
which is determined by the chiral unitary approach, the Pauli spinor
$\chi _{Y}$ for the hyperon, and $K^{-}$, $\pi$, and hyperon momenta
$p_{K^{-}}$, $p_{\pi}$, and $p_{Y}$, respectively.  By means of the
two summation symbols with $\lambda$ and $\lambda ^{\prime}$, the sum
and average of the scattering amplitude for the polarizations of
baryons are done, and $(\pi , Y)$ under the summation symbol
represents the absorption channels to $\pi \Sigma$ and $\pi \Lambda$.
Performing the integrations in Eqs.~\eqref{eq:U-one} and
\eqref{eq:gamma-one}, we obtain,
\begin{equation} 
2 \text{Im} U_{K^{-}}^{\text{one}} 
= - 
\int
_{0}^{k_{\text{F}}} \frac{d p_{1} \, p_{1}^{2}}{\pi ^{2}} 
\overline{\sum_{\lambda}} \sum _{\lambda ^{\prime}} 
\sum _{(\pi , Y)}
\gamma _{\pi
  Y} (\mu _{1} ; \, k_{\text{F}}) ,
\label{eq:U-one2}
\end{equation}
\begin{equation}
\gamma _{\pi Y} (\mu _{1} ; \, k_{\text{F}}) 
=  
\frac{p_{\text{cm}}^{\prime} M_{Y}}{8 \pi ^{2} W} 
\int d \Omega _{Y} 
\left | 
\chi _{Y}^{\dagger} 
{\cal T}_{\pi Y} ( W ) 
\chi _{1} 
\right | ^{2} , 
\label{eq:gamma-one2}
\end{equation}
with the center-of-mass energy of $K^{-} N_{1}$ system $W$, 
\begin{align}
W 
& = \sqrt{(E_{1} + E_{K^{-}})^{2} - (\bm{p}_{1} + \bm{p}_{K^{-}})^{2}} 
\end{align}
and the initial nucleon energy $E_{1}$ expressed in
Eq.~\eqref{eq:ThomasFermi}, and the momentum of the center-of-mass
frame for the on-shell $\pi Y$ state $p_{\text{cm}}^{\prime}$.
Here we will take an angular average for the kaon momentum 
in the integral of the nucleon momentum as
\begin{align}
W 
& \approx \sqrt{(E_{1} + E_{K^{-}})^{2} - (p_{1}^{2}+ p_{K^{-}}^{2})} .
\label{eq:W-one}
\end{align}

Next let us consider the nonmesonic absorption process.  Taking into
account the Feynman diagrams shown in Fig.~\ref{fig:Feynman-K2}, the
cut amplitude for the nonmesonic process can be written as,
\begin{align}
2 \text{Im} U_{K^{-}}^{\text{two}} 
= & -  
\int ^{k_{\text{F}}} \frac{d^{3} p_{1}}{(2 \pi)^{3}} g_{N} 
\int ^{k_{\text{F}}} \frac{d^{3} p_{2}}{(2 \pi)^{3}} g_{N} 
\nonumber \\ 
& \times \overline{\sum_{\lambda}} \sum _{\lambda ^{\prime}} 
\sum _{(Y, N)}
\gamma _{Y N} (\mu _{1}, \, \mu _{2} ; \, k_{\text{F}}) , 
\label{eq:U-two}
\end{align}
with the reaction rate for the $K^{-} NN \to Y N$ process 
$\gamma _{YN}$ defined as, 
\begin{align}
& \gamma _{Y N} (\mu _{1}, \, \mu _{2}  ; \, k_{\text{F}}) 
\equiv 
\int d \Phi _{Y N} 
\left | {\cal A}_{YN} \right | ^{2} \eta _{YN}
\nonumber \\ & \times 
(2 \pi)^{4} \delta ^{4} (p_{K^{-}} + p_{1} + p_{2} - p_{Y} - p_{N}) , 
\label{eq:gamma-two}
\end{align}
with, 
\begin{equation}
d \Phi _{Y N} \equiv 
\frac{d^{3} p_{Y}}{(2 \pi)^{3}} \frac{2 M_{Y}}{2 E_{Y}} 
\frac{d^{3} p_{N}}{(2 \pi)^{3}} \frac{2 M_{N}}{2 E_{N}} . 
\end{equation}
Here ${\cal A}_{YN}$ is the scattering amplitude for the $K^{-} NN \to
Y N$ process, and $p_{Y}$ and $p_{N}$ are the hyperon and nucleon
momenta in the final state, respectively.  The symbol $\eta _{YN}$ is
defined to be,
\begin{align}
\eta _{YN} = 
\begin{cases}
& 2 \quad \text{for } Y N 
= \Lambda n, \, \Sigma ^{0} n , \text{ and } \Sigma ^{-}p, \\
& 1 \quad \text{for others,}
\end{cases}
\end{align}
in order to take into account the same contribution from initial $pn$
state with exchanged quantum numbers, namely $p (\mu _{1}) n (\mu
_{2})$ and $n (\mu _{1}) p (\mu _{2})$ for $K^{-} pn \to \Lambda n$,
$\Sigma ^{0} n$, and $\Sigma ^{-} p$ reactions.  The scattering
amplitude ${\cal A}_{YN}$ can be written by summing all
possible channels labeled by $i$ as,

\begin{table}[t]
  \caption{\label{tab:Feynman-b} 
    Possible channels for Eqs.~\eqref{eq:Amp-two1}--\eqref{eq:Amp-two3}.
    Here $N_{1}$ and $N_{2}$ are the nucleons in the initial state, $a$ is
    the exchange meson, and $A$ and $B$ are the baryons in the final
    state.  $\xi$ is the relative sign of the amplitude coming from the
    exchange of the final state baryons, and $\alpha$ and $\beta$ are the
    Clebsch-Gordan coefficients for the $MBB$ coupling. }
  \begin{ruledtabular}
    \begin{tabular}{cccccccc} 
      $N_{1}$ & $N_{2}$ & $a$ & $A$ & $B$ & $\xi$ & $\alpha$ & $\beta$ \\ 
      \hline
      $p$ & $p$ & 
      \rule[0pt]{0pt}{9pt} $K ^{-}$ & $p$ & $\Lambda$ & 
      $+$ & $-2/\sqrt{3}$ & $1/\sqrt{3}$ \\
      & & $\eta$ & $\Lambda$ & $p$ & 
      $-$ & $1/\sqrt{3}$ & $-2/\sqrt{3}$ \\
      & & $\pi ^{0}$ & $\Lambda$ & $p$ & 
      $-$ & $1$ & $0$ \\
      \hline
      $p$ & $p$ &       \rule[0pt]{0pt}{9pt} 
      $K ^{-}$ & $p$ & $\Sigma ^{0}$ & 
      $+$ & $0$ & $1$ \\
      & & $\pi ^{0}$ & $\Sigma ^{0}$ & $p$ & 
      $-$ & $1$ & $0$ \\
      & & $\eta$ & $\Sigma ^{0}$ & $p$ & 
      $-$ & $1/\sqrt{3}$ & $-2/\sqrt{3}$ \\
      \hline
      $p$ & $p$ & \rule[0pt]{0pt}{9pt} 
      $\bar{K} ^{0}$ & $n$ & $\Sigma ^{+}$ & 
      $+$ & $0$ & $\sqrt{2}$ \\
      & & $\pi ^{-}$ & $\Sigma ^{+}$ & $n$ & 
      $-$ & $\sqrt{2}$ & $0$ \\
      \hline \hline 
      $p$ & $n$ & \rule[0pt]{0pt}{9pt} 
      $\bar{K} ^{0}$ & $n$ & $\Lambda$ & 
      $+$ & $-2/\sqrt{3}$ & $1/\sqrt{3}$ \\
      & & $\eta$ & $\Lambda$ & $n$ & 
      $-$ & $1/\sqrt{3}$ & $-2/\sqrt{3}$ \\
      & & $\pi ^{0}$ & $\Lambda$ & $n$ & 
      $-$ & $-1$ & $0$ \\
      $n$ & $p$ & 
      $K ^{-}$ & $n$ & $\Lambda$ & 
      $+$ & $-2/\sqrt{3}$ & $1/\sqrt{3}$ \\
      & & $\pi ^{-}$ & $\Lambda$ & $n$ & 
      $-$ & $\sqrt{2}$ & $0$ \\
      \hline 
      $p$ & $n$ & \rule[0pt]{0pt}{9pt} 
      $\bar{K} ^{0}$ & $n$ & $\Sigma ^{0}$ & 
      $+$ & $0$ & $-1$ \\
      & & $\pi ^{0}$ & $\Sigma ^{0}$ & $n$ & 
      $-$ & $-1$ & $0$ \\
      & & $\eta$ & $\Sigma ^{0}$ & $n$ & 
      $-$ & $1/\sqrt{3}$ & $-2/\sqrt{3}$ \\
      $n$ & $p$ & 
      $K ^{-}$ & $n$ & $\Sigma ^{0}$ & 
      $+$ & $0$ & $1$ \\
      & & $\pi ^{-}$ & $\Sigma ^{0}$ & $n$ & 
      $-$ & $\sqrt{2}$ & $0$ \\
      \hline 
      $p$ & $n$ & $K^{-}$ & $p$ & $\Sigma ^{-}$ & 
      $+$ & $0$ & $\sqrt{2}$ \\
      & & $\pi ^{+}$ & $\Sigma ^{-}$ & $p$ & 
      $-$ & $\sqrt{2}$ & $0$ \\
      $n$ & $p$ & 
      $\pi ^{0}$ & $\Sigma ^{-}$ & $p$ & 
      $-$ & $1$ & $0$ \\
      & & $\eta$ & $\Sigma ^{-}$ & $p$ & 
      $-$ & $1/\sqrt{3}$ & $-2/\sqrt{3}$ \\
      \hline \hline
      $n$ & $n$ & \rule[0pt]{0pt}{9pt} $K ^{-}$ & $n$ & $\Sigma ^{-}$ & 
      $+$ & $0$ & $\sqrt{2}$ \\
      & & $\pi ^{0}$ & $\Sigma ^{-}$ & $n$ & 
      $-$ & $-1$ & $0$ \\
      & & $\eta$ & $\Sigma ^{-}$ & $n$ & 
      $-$ & $1/\sqrt{3}$ & $-2/\sqrt{3}$ \\
    \end{tabular}
  \end{ruledtabular}
\end{table}%

\begin{align}
{\cal A}_{\Lambda p, \, \Sigma ^{0} p, \, \Sigma ^{+} n} = 
& \frac{1}{\sqrt{2}} \sum _{i} \xi _{i}
\Bigg [
{\cal A}_{i} (K^{-} p (\mu _{1}) p (\mu _{2}) 
\xrightarrow{a_{i}} A_{i} B_{i}) 
\nonumber \\
& - {\cal A}_{i} (K^{-} p (\mu _{2}) p (\mu _{1}) 
\xrightarrow{a_{i}} A_{i} B_{i}) 
\Bigg ] , 
\label{eq:Amp-two1}
\end{align}
for the $K^{-}pp \to \Lambda p$, $\Sigma ^{0} p$, and $\Sigma ^{+} n$
reactions, 
\begin{align}
& {\cal A}_{\Lambda n, \, \Sigma ^{0} n, \, \Sigma ^{-} p} 
= 
\frac{1}{\sqrt{2}} \Bigg [ 
\sum _{i} \xi _{i}
{\cal A}_{i} (K^{-} p (\mu _{1}) n (\mu _{2}) 
\xrightarrow{a_{i}} A_{i} B_{i}) 
\nonumber \\ 
& - \sum _{i} \xi _{i}
{\cal A}_{i} (K^{-} n (\mu _{2}) p (\mu _{1}) 
\xrightarrow{a_{i}} A_{i} B_{i}) \Bigg ], 
\label{eq:Amp-two2}
\end{align}
for the $K^{-}pn \to \Lambda n$, $\Sigma ^{0} n$, and $\Sigma ^{-} p$
reactions, and 
\begin{align}
{\cal A}_{\Sigma ^{-} n} = 
& \frac{1}{\sqrt{2}} \sum _{i} \xi _{i} \Bigg [
{\cal A}_{i} (K^{-} n (\mu _{1}) n (\mu _{2}) 
\xrightarrow{a_{i}} A_{i} B_{i}) 
\nonumber \\
& - {\cal A}_{i} (K^{-} n (\mu _{2}) n (\mu _{1}) 
\xrightarrow{a_{i}} A_{i} B_{i}) 
\Bigg ] , 
\label{eq:Amp-two3}
\end{align}
for the $K^{-}nn \to \Sigma ^{-} n$ reaction.  Here $a_{i}$ represents
the exchange meson, and $A_{i}$ and $B_{i}$ are the baryons in the
final state. The explicit channels are given in
Table~\ref{tab:Feynman-b}.  The amplitude ${\cal A}$ for the $K^{-}
N_{1}(\mu _{1}) N_{2}(\mu _{2}) \xrightarrow{a} A B$ process is
calculated in the one-meson exchange model,
\begin{align}
& {\cal A}_{i} ( K^{-} N_{1} (\mu _{1}) N_{2} (\mu _{2}) \xrightarrow{a} A B) 
\nonumber \\
& = 
\chi _{A}^{\dagger} 
{\cal T}_{a A} (W) 
\chi _{1} 
\times 
\tilde{\Pi}_{a} (q_{a}^{2})
\times 
\tilde{V}_{aN_{2}B} 
\chi _{B}^{\dagger} 
\bm{q}_{a} \cdot \bm{\sigma}
\chi _{2} 
, 
\label{eq:Amp_each}
\end{align}
where the symbol $\xi$ denotes the relative sign of the amplitude
coming from the exchange of the final-state baryons.  In the amplitude
${\cal A}_{i}$, ${\cal T}_{aA}(W)$ is the $K^{-}N_{1}\to a A$
scattering amplitude, which is determined by the chiral unitary
approach, with the energy $W$ expressed in Eq.~\eqref{eq:W-one}.  The
meson propagator $\tilde{\Pi}_{a}$ with the meson momentum
$q^{\mu}=p_{B}^{\mu} - p_{2}^{\mu}$ includes the short-range
correlations~\cite{Oset:1979bi},
\be
\tilde{\Pi}_{a} (q^{2}) = \left ( \frac{\Lambda ^{2}}{\Lambda ^{2} -
    q^{2}} \right ) ^{2} \frac{1}{q^{2} - m_{a}^{2}} - \left (
  \frac{\Lambda ^{2}}{\Lambda ^{2} - \tilde{q}^{2}} \right ) ^{2}
\frac{1}{\tilde{q}^{2} - m_{a}^{2}} , 
\ee 
with $\tilde{q}^{2} = q^{2}
- q_{\text{C}}^{2}$, where we choose typical parameter set, $\Lambda =
1.0 \gev$ and $q_{\text{C}}=780 \mev$~\cite{Jido:2001am}.  The coefficient
of the meson-baryon-baryon coupling $\tilde{V}_{a N_{2}B}$ is
determined by the flavor SU(3) symmetry as, 
\be \tilde{V}_{a N_{2} B}
= \alpha _{a N_{2} B} \frac{D + F}{2 f} + \beta _{a N_{2} B} \frac{D -
  F}{2 f} , 
\ee 
with empirical values of $D+F=1.26$ and $D-F=0.33$, which reproduce
the hyperon $\beta$ decays observed in experiments, and
$f=f_{\pi}=93.0 \mev$ commonly for all the mesons.  The SU(3)
Clebsch-Gordan coefficients $\alpha$ and $\beta$ are listed in
Table~\ref{tab:Feynman-b}.  The $K^{-}N_{1}\to a A$ scattering
amplitude ${\cal T}_{aA}$ has the indices of spinors for $N_{1}$
($\chi _{1}$) and $A$ ($\chi _{A}$), whereas the Pauli matrices
$\sigma ^{i}$ ($i=1,\, 2,\, 3$) appearing in Eq.~\eqref{eq:Amp_each}
are given in the space of the spinors for $N_{2}$ ($\chi _{2}$) and
$B$ ($\chi _{B}$).  We emphasize that the antisymmetric combinations
for initial nucleons are realized as, {\it e.g.}, $(|p (\mu _{1}) p
(\mu _{2}) \rangle - |p (\mu _{2}) p (\mu _{1}) \rangle)/\sqrt{2}$ in
the amplitudes ${\cal A}_{\Lambda p, \, \Sigma ^{0} p, \, \Sigma ^{+}
  n}$ [see Eqs.~\eqref{eq:Amp-two1}--\eqref{eq:Amp-two3} and
Table~\ref{tab:Feynman-b}].

Performing the integrations in Eqs.~\eqref{eq:U-two} and
\eqref{eq:gamma-two}, we obtain,
\begin{align}
2 \text{Im} U_{K^{-}}^{\text{two}} 
= - &
\int _{0}^{k_{\text{F}}} \frac{d p_{1} \, p_{1}^{2}}{\pi ^{2}} 
\int _{0}^{k_{\text{F}}} \frac{d p_{2} \, p_{2}^{2}}{\pi ^{2}} 
\int _{-1}^{1} \frac{d \cos \theta _{12}}{2} 
\nonumber \\
& \times 
\overline{\sum_{\lambda}} \sum _{\lambda ^{\prime}} 
\sum _{(Y, N)}
\gamma _{Y N} (\mu _{1}, \, \mu _{2}; \, k_{\text{F}}) , 
\end{align}
\begin{align}
& \gamma _{Y N} (\mu _{1}, \, \mu _{2}; \, k_{\text{F}}) 
=  
\frac{p_{\text{cm}}^{\prime \prime} M_{Y} M_{N}}{4 \pi ^{2} E_{\text{tot}}} 
\int d \Omega _{N} 
\left | 
{\cal A} _{Y N} 
\right | ^{2} \eta _{YN} , 
\end{align}
with momentum $p_{\text{cm}}^{\prime \prime}$ for the on-shell $Y$ and
$N$ states in the center-of-mass frame and total energy
$E_{\text{tot}} = \sqrt{(p_{Y}+p_{N})^{2}}$.

\section{Results}
\label{sec:3}

We now show our results for the $K^{-}$ absorption potential as a
function of nuclear density $\rho _{N}$.  First we consider the
self-energy of kaon at rest in nuclear matter with
$p_{K^{-}}^{\mu}=(m_{\bar{K}}, \, \bm{0})$. Next we see the absorption
widths for kaons with finite momenta and energies in
Sec.~\ref{sec:FiniteEP}.  In this section we concentrate on
contributions from the $s$-wave $\KbarN$ interaction in the $K^{-}$
absorption reaction, because, as we have already mentioned, the
resonance $\LamFOF$ ($\LamStar$) just below the $\KbarN$ threshold in
$s$ wave will play the most important role in the $K^{-}$ absorption
process.  Later we will discuss the $\SigTEF$ contributions in
Sec.~\ref{sec:Sigma1385}.

\begin{figure}[!Ht]
  \centering
  \Psfig{8.6cm}{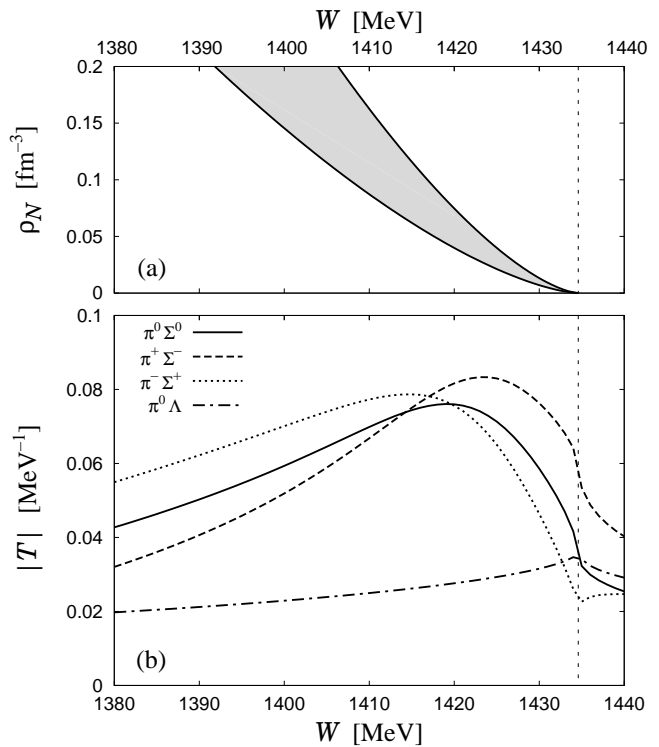}
  \caption{\label{fig:W_region} Range of energy which can be achieved
    by a fixed energy $W$ of the $K^{-}$ at rest and a bound nucleon
    in the nucleus [the shaded area of (a)].  We also show absolute
    values of the scattering amplitude $|{\cal T}|$ for the $K^{-}p
    \to (\pi \Sigma )^{0}$ and $\pi ^{0} \Lambda$ transitions as
    functions of the energy of the $K^{-}p$ system, $W$ (b).  The
    vertical lines represent the $\KbarN$ threshold. }
\end{figure}

First of all, let us recall that the energy of a two-body system of
the kaon at rest and a nucleon inside the nucleus can be less than the
$\KbarN$ threshold energy owing to the off-shellness of the bound
nucleons. The accessible energy range depends on the Fermi momentum
for the nucleons, namely the nuclear density, as shown in
Eq.~\eqref{eq:ThomasFermi}.  The relation between the accessible
energy range and the nuclear density is shown in
Fig.~\ref{fig:W_region}(a).  As one can see from
Fig.~\ref{fig:W_region}(a), $\bar{K}$-$N$ two body systems can have
lower energies in higher densities and vice versa.  Oppositely, there
is a range of density in which a fixed value $W$ can be achieved by
the energy of $K^{-} N$, as shown in Fig.~\ref{fig:W_region}(a).  This
means that strength of the $\LamStar$ contribution to the $K^{-}p \to
MB$ transitions in absorption reactions depends on the nuclear
density.  Hence, in order to see in which density $\LamStar$ appears
in the absorption reaction, we show the absolute values of the
scattering amplitude for the $K^{-}p \to (\pi \Sigma )^{0}$ and $\pi
^{0} \Lambda$ transitions in Fig.~\ref{fig:W_region}(b).  From
Fig.~\ref{fig:W_region} we can see that the $\LamStar$ spectra in the
$(\pi \Sigma)^{0}$ channels have a peak around 1420 MeV with a 40 MeV
width, which energy can be achieved by a pair of $K^{-} p$ in nuclear
matter with the density $\rho _{N} \approx 0.05$--$0.1 \fm ^{-3}$.  We
also see that at the saturation density $\rho _{0}=0.17 \fm^{-3}$ the
energy of the $K^{-} p$ pair is around $1400 \mev$, which is in the
$\LamStar$ resonance peak.  Owing to the presence of the $\LamStar$
resonance, the amplitudes have strong energy dependence.  This will
make nontrivial $\rho_{N}$ dependence to mesonic and nonmesonic
absorption potentials.

It is also important noting that the peak structure in the $(\pi
\Sigma )^{0}$ amplitude comes from $\LamStar$ with $I=0$ but the peak
position is slightly different in each charged channel. This is
because the $I=1$ nonresonant contributions are not so small and the
interference between the $I=0$ and $I=1$ contributes in the opposite
way for the $\pi^{\pm} \Sigma^{\mp}$ channels.

\subsection{Mesonic absorption}
\label{sec:one-body}

First we consider the mesonic absorption potential of $K^{-}$.  We
note that in mesonic absorption $K^{-}p \to (\pi \Sigma )^{0}$
processes contain the $\LamStar$ resonance whereas $K^{-}p \to \pi
^{0} \Lambda$ and $K^{-}n \to (\pi Y)^{-}$ processes do not have the
$\LamStar$ contributions.  We also note that we expect that the
mesonic absorption potential would be proportional to $\rho _{N}$, if
the $\KbarN$ amplitude would not depend on energy.

\begin{figure}[!Ht]
  \centering
    \Psfig{8.6cm}{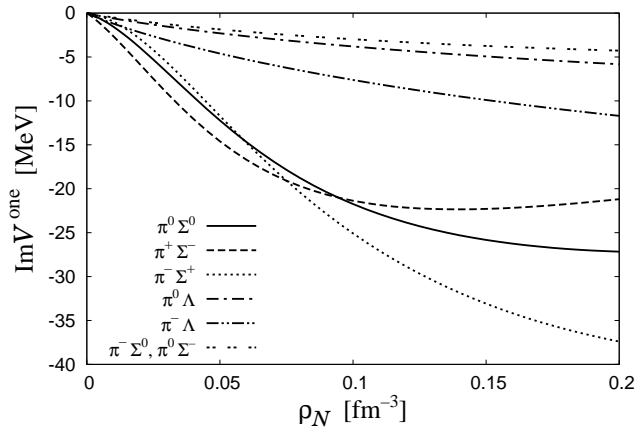}
    \caption{Mesonic absorption potential ($\text{Im} V^{\text{one}}$)
      for $K^{-}$ at rest in nuclear matter as a function of nuclear
      density.  The potentials for $K^{-} n \to \pi ^{-} \Sigma ^{0}$
      and $\pi ^{0} \Sigma ^{-}$ have the same values owing to the
      isospin symmetry.  }
  \label{fig:abs-one}
\end{figure}

In Fig.~\ref{fig:abs-one}, we show the result of the mesonic
absorption potential of $K^{-}$ at rest in nuclear matter. From the
figure, we find that absorption to the $(\pi \Sigma)^{0}$ states
are dominant to the other channels.  Since the $\LamStar$ resonance
appears selectively in the $K^{-} p \to (\pi \Sigma)^{0}$ transitions,
this result shows that the $\LamStar$ contribution is indeed important
for the mesonic absorption of $K^{-}$ in these densities, and that
$K^{-}$ at rest is absorbed through the $\LamStar$ resonance
($\LamStar$ doorway process).  Thus, if one observes large branching
ratios of $(\pi \Sigma)^{0}$ in $K^{-}$ absorption into nuclei, this
observation indicates that the $\LamStar$ doorway process dominates
the $K^{-}$ absorption reaction.  As for the density dependence of the
mesonic absorption potential, the potential for the $(\pi \Sigma
)^{0}$ channels do not show $\rho _{N} ^{1}$-like dependence around
$\rho _{N} > 0.1 \fm ^{-3} \approx 0.6 \rho _{0}$ whereas that for the
$\pi ^{0} \Lambda$, $\pi ^{-} \Lambda$, and $(\pi \Sigma )^{-}$ states
shows $\rho _{N}^{1}$ dependence.  This is owing to the energy
dependence of the $\KbarN$ amplitude coming from the $\LamStar$
resonance.

\begin{figure}[!Ht]
  \centering
    \Psfig{8.6cm}{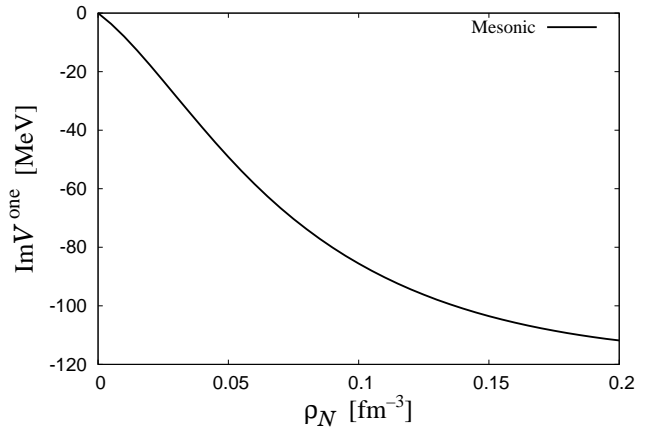}
    \caption{Total sum of mesonic absorption potential ($\text{Im}
      V^{\text{one}}$) for $K^{-}$ at rest in nuclear matter as a
      function of nuclear density.  }
  \label{fig:abs-one-total}
\end{figure}

\begin{figure}[!Ht]
  \centering
    \Psfig{8.6cm}{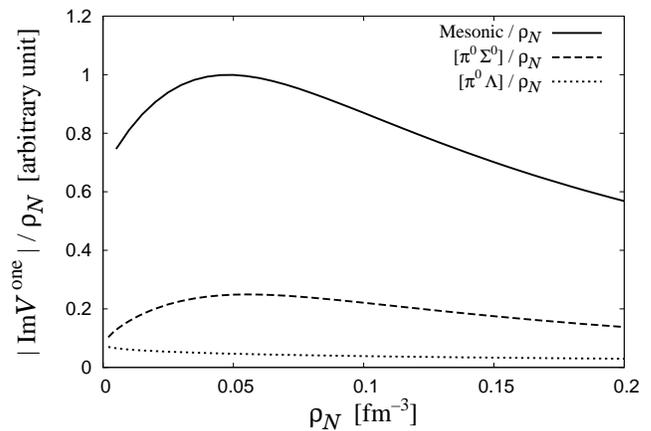}
    \caption{Absolute absorption potentials divided by $\rho _{N}$ for
      total, $\pi ^{0} \Sigma ^{0}$, and $\pi ^{0} \Lambda$ contributions
      as functions of nuclear density in arbitrary unit. }
  \label{fig:abs-one-rho}
\end{figure}

\begin{figure}[!Ht]
  \centering
    \Psfig{8.6cm}{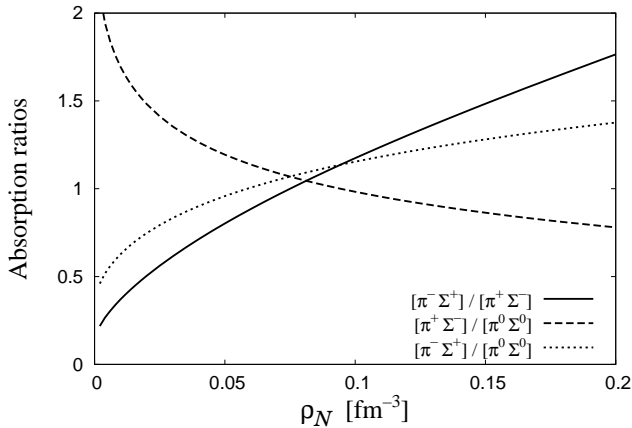}
    \caption{Ratios of mesonic absorption potentials for $[\pi ^{-}
      \Sigma ^{+}] / [\pi ^{+} \Sigma ^{-}]$, $[\pi ^{+} \Sigma ^{-}]
      / [\pi ^{0} \Sigma ^{0}]$, and $[\pi ^{-} \Sigma ^{+}] / [\pi
      ^{0} \Sigma ^{0}]$ as functions of nuclear density.  }
  \label{fig:ratio-one}
\end{figure}

\begin{figure*}[!t]
  \centering
  \begin{tabular*}{\textwidth}{@{\extracolsep{\fill}}cc}
    \Psfig{8.6cm}{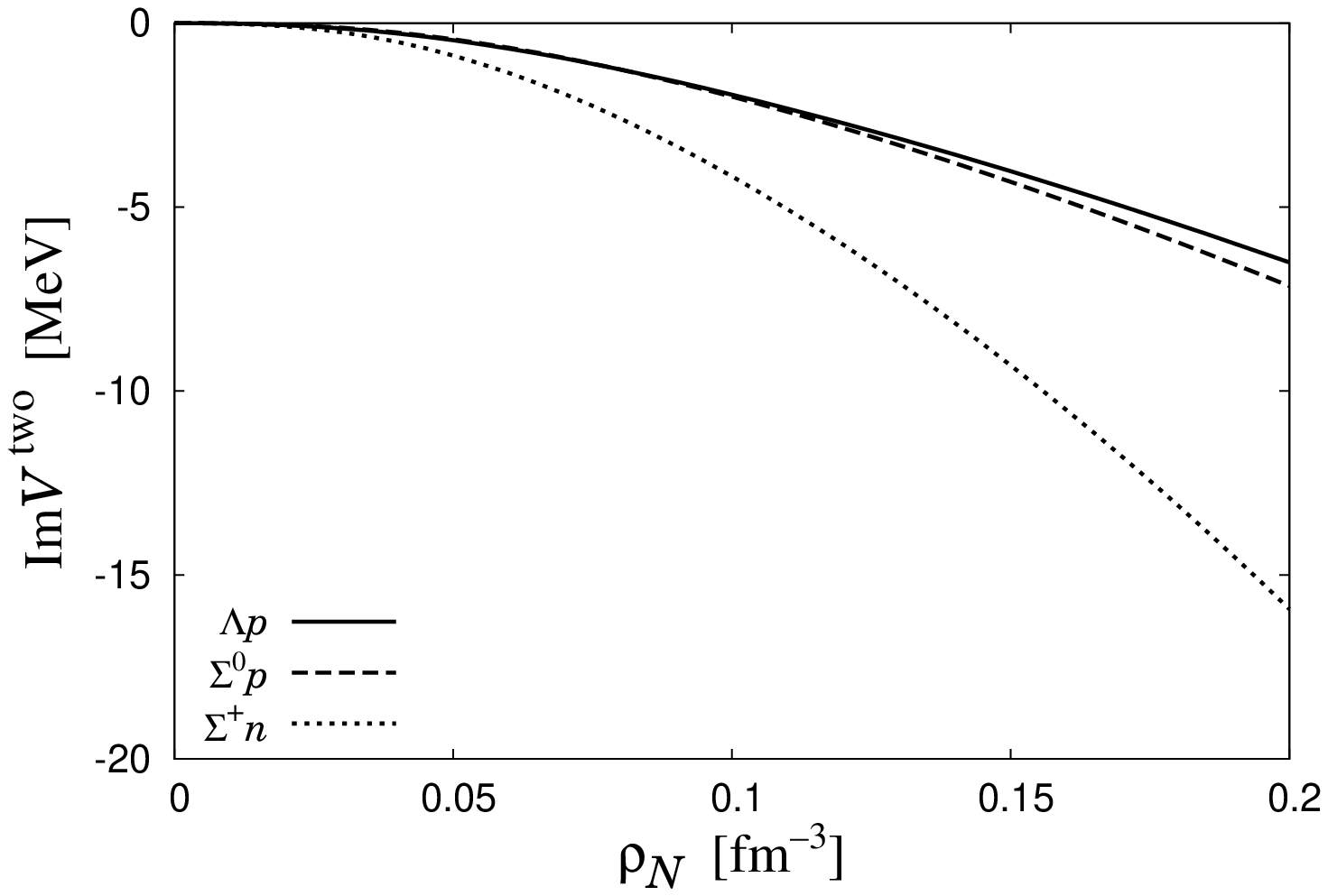} & 
    \Psfig{8.6cm}{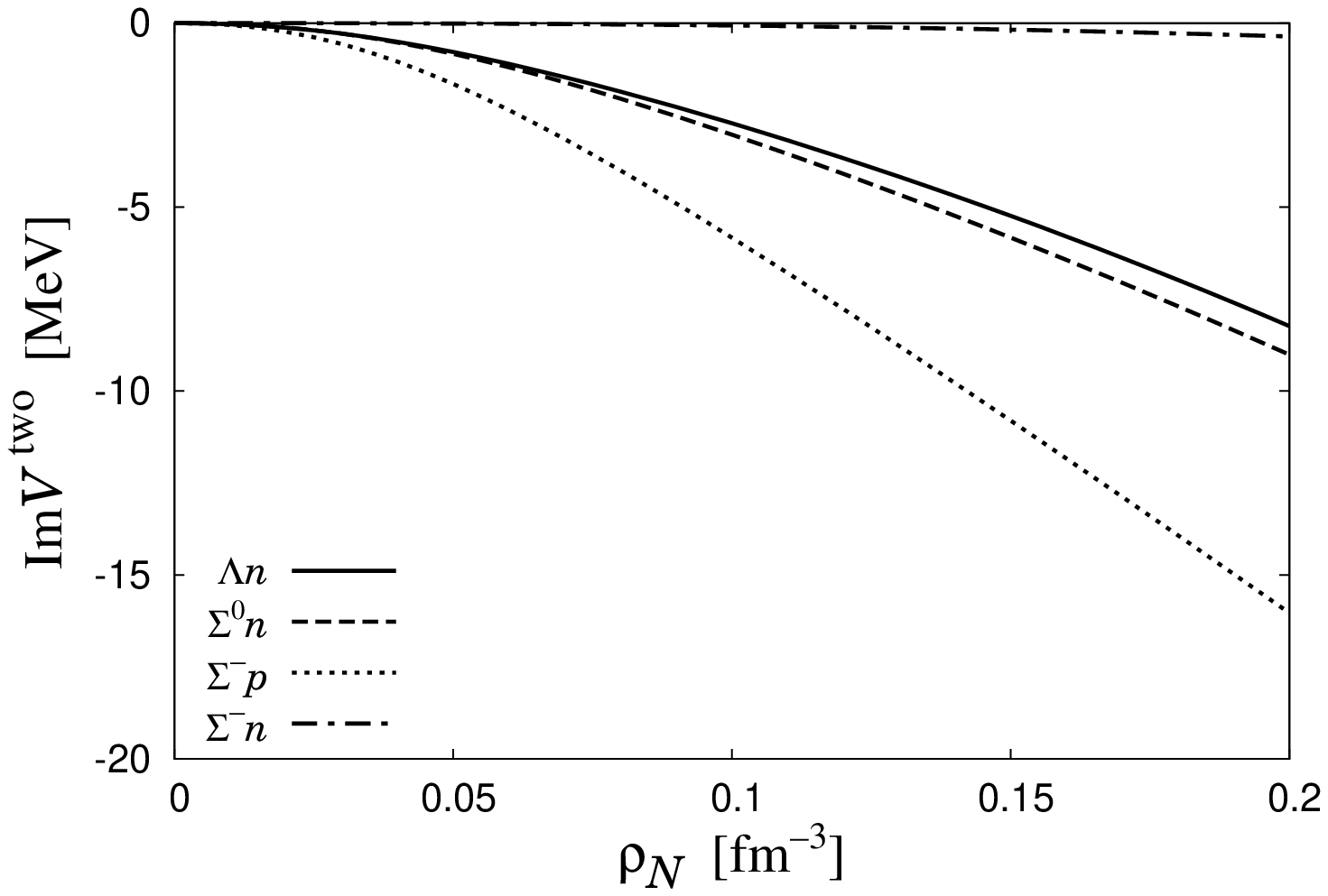} \\
    (a) & (b)
  \end{tabular*}
  \caption{Nonmesonic absorption potential ($\text{Im}
    V^{\text{two}}$) for $K^{-}$ at rest in nuclear matter as a
    function of nuclear density.  The contributions of the $K^{-}pp
    \to \Lambda p$, $\Sigma ^{0} p$, and $\Sigma ^{+} n$ processes (a)
    and of the $K^{-}pn \to \Lambda n$, $\Sigma ^{0} n$, $\Sigma ^{-}
    p$ and $K^{-}nn \to \Sigma ^{-} n$ processes (b). }
  \label{fig:abs-two} 
\end{figure*}%

The total sum of the mesonic absorption potential is shown in
Fig.~\ref{fig:abs-one-total} as a function of nuclear density.  The
total value of the mesonic absorption width ($=-2
\text{Im}V^{\text{one}}$) amounts to about $200 \mev$ at the
saturation density ($\rho _{0}= 0.17 \fm ^{-3}$).  The large value of
the absorption width is caused because, in addition to that the
$K^{-}p$ energy in the realistic nuclear density is within the range
of the $\LamStar$ peak, the number of the initial-state nucleons which
can create $\LamStar$ becomes large as the nuclear density increases,
as seen in Eq.~\eqref{eq:U-one2}.  Here we note that moderate
absorption width will be obtained when in-medium $\KbarN$ scattering
amplitude rather than the free space is used.  Indeed, by using an
approximation,
\begin{equation}
\sum _{(\pi , Y)}
\frac{p_{\rm cm}^{\prime} M_{Y}}{8 \pi ^{2} W} 
\int d \Omega _{Y} |{\cal T}_{\pi Y}^{\rm med} ( W; \, \rho ) | 
\approx - 2 \text{Im} T_{K^{-} p}^{\rm med} ( W; \, \rho ) , 
\end{equation}
for the in-medium $K^{-} N \to MB$ scattering amplitude $T_{MB}^{\rm
  med} ( W; \, \rho )$ and taking value of $\text{Im} T_{K^{-} p}^{\rm
  med} ( W; \, \rho )$ from Ref.~\cite{Ramos:1999ku}, we roughly
estimate the mesonic absorption potential with the in-medium amplitude
to be $\text{Im} V^{\text{one}} \sim -40 \mev$ at the saturation
density $\rho _{0}$.  The obtained value is about two-fifths of our
results (see Fig.~\ref{fig:abs-one-total}) and consistent with thse
preceding works~\cite{Ramos:1999ku, Cieply:2011fy}, in which in-medium
scattering amplitudes are employed to calculate the absorption
potential.  The total mesonic absorption potential shows non-$\rho
_{N}^{1}$ dependence because of the $\LamStar$ doorway contributions.
In order to see at which density the $\LamStar$ contribution is large,
we plot in Fig.~\ref{fig:abs-one-rho} the absolute absorption
potentials divided by the nuclear density,
$|\text{Im}V^{\text{one}}|/\rho _{N}$, for the total and the $\pi ^{0}
\Sigma ^{0}$ and $\pi ^{0} \Lambda$ mesonic channels as functions of
the nuclear density.  As one can see from Eqs.~\eqref{eq:U-one2} and
\eqref{eq:gamma-one2}, $|\text{Im}V^{\text{one}}|/\rho _{N}$ takes
value approximately proportional to the squared scattering amplitude
$|{\cal T}_{\pi Y}|^{2}$ with energies achieved by the considering
nuclear density [see also Fig.~\ref{fig:W_region}(a)].  Therefore,
$|\text{Im}V^{\text{one}}|/\rho _{N}$ reflects the structure of the
$K^{-} N \to \pi Y$ transition process.  From
Fig.~\ref{fig:abs-one-rho}, there is no structure in the potential for
the $\pi ^{0} \Lambda$ channel divided by $\rho _{N}$, because the
$K^{-}p \to \pi ^{0} \Lambda$ process does not have the $\LamStar$
contribution.  On the other hand, a local maximum appears at $\rho
_{N} \approx 0.05 \fm ^{-3} \approx 0.3 \rho _{0}$ in case of the
total as well as the $\pi ^{0} \Sigma ^{0}$ channel, which indicates
enhancement of absorption, due to the $\LamStar$ doorway contribution.
The position of the maximum reflects the matching condition of
$K^{-}p$ energy $W$ to the $\LamStar$ peak position via the $\LamStar$
resonance contribution.  The fact that a local maximum of
$|\text{Im}V^{\text{one}}|/\rho _{N}$ appears at $\rho _{N} \approx
0.05 \fm ^{-3} \approx 0.3 \rho _{0}$ is expected from
Fig.~\ref{fig:W_region}, which shows that this nuclear density
corresponds to the peak position of $\LamStar$ in $K^{-}p$ energy $W
\approx 1420 \mev$.

Another interesting feature of the absorption potential shown in
Fig.~\ref{fig:abs-one} is that the behavior of the absorption to the
$\pi ^{0} \Sigma ^{0}$, $\pi ^{+} \Sigma ^{-}$, and $\pi ^{-} \Sigma
^{+}$ channels is different from each other, especially at higher
densities ($\gtrsim 0.1 \fm ^{-3} \approx 0.6 \rho _{0}$).  This comes
from the slight difference of the $\LamStar$ spectrum in each channel
stemming from the interference between $\LamStar$ in $I=0$ and the
$I=1$ nonresonant background as shown in Fig.~\ref{fig:W_region}(b).
At the saturation density the $K^{-} p$ energy achieves $\approx 1400
\mev$, where the squared amplitude $|{\cal T}_{\pi ^{-} \Sigma
  ^{+}}|^{2}$ is about two times larger than $|{\cal T}_{\pi ^{+}
  \Sigma ^{-}}|^{2}$ (see Fig.~\ref{fig:W_region}), hence the
absorption to the $\pi ^{-} \Sigma ^{+}$ channel becomes about two
times larger than the $\pi ^{+} \Sigma ^{-}$ channel.

Even though the $\LamStar$ resonance sits in the $I=0$ channel, the
interference between the $I=0$ and $I=1$ contributions makes the peak
position of the $\LamStar$ spectrum shift in the opposite direction in
the $\pi^{\pm} \Sigma^{\mp}$ channels as seen in
Fig.~\ref{fig:W_region}. The effect of the peak shift can be clearly
seen in the density dependence of the ratios of $K^{-}$ absorption
into $(\pi \Sigma)^{0}$ channels, because the density determines the
accessible energy of the two-body system of $K^{-}$ and a bound
nucleon.  In Fig.~\ref{fig:ratio-one} we plot the ratios of the
mesonic absorption potential for the $\pi ^{0} \Sigma ^{0}$, $\pi ^{+}
\Sigma ^{-}$, and $\pi ^{-} \Sigma ^{+}$ channels.  As one can see,
while the ratio $[\pi ^{-} \Sigma ^{+}] / [\pi ^{+} \Sigma ^{-}]$,
which we denote $R_{+-}$, is less than unity in $\rho _{N} < 0.08 \fm
^{-3} \approx 0.5 \rho _{0}$, it gets larger as the density increases
and becomes $\sim 1.6$ at the saturation density.  This tendency comes
from the facts that the upward shift of the $\LamStar$ peak is seen in
the $K^{-}p \to \pi^{+} \Sigma^{-}$ amplitude while the downward shift
in $K^{-}p \to \pi^{-} \Sigma^{+}$ and that the smaller Fermi momentum
for the nucleon, or the lower density, probes the $\LamStar$ spectrum
in energies closer to the threshold, while the higher density probes
the lower energy of the $\LamStar$ spectrum.  We also show the ratios
of the mesonic absorption potentials for $[\pi^{\pm} \Sigma^{\mp}] /
[\pi^{0} \Sigma^{0}]$ in Fig.~\ref{fig:ratio-one}.  Here we note that
the $\pi^{0} \Sigma^{0}$ channel has no $I=1$ contribution and can be
a guide for the $\LamStar$ spectrum.  The ratio $[\pi ^{\pm} \Sigma
^{\mp}] / [\pi ^{0} \Sigma ^{0}]$ shows opposite behaviors to each
other; $[\pi ^{+} \Sigma ^{-}] / [\pi ^{0} \Sigma ^{0}]$ ($[\pi ^{-}
\Sigma ^{+}] / [\pi ^{0} \Sigma ^{0}]$) becomes weaker (large) as the
density increases.  All of the three ratios in
Fig.~\ref{fig:ratio-one} is almost unity at $\rho _{N}\approx 0.08 \fm
^{-3} \approx 0.5 \rho _{0}$.

The increase of the absorption ratio $R_{+-}$ as the density increases
also indicates the nature of the $\LamStar$ resonance.  As mentioned
before, the increase of the ratio means that the $\LamStar$ peak is
shifted upward in the $\pi^{+} \Sigma^{-}$ channel and downward in the
$\pi^{-} \Sigma^{+}$ channel, which is a consequence of the
interference of $I=0$ and $I=1$ and is determined by the relative sign
of the $I=0$ and $I=1$ amplitudes.  Then, an important point is that
the inversion of the $\pi^{+} \Sigma^{-}$ dominance to the $\pi^{-}
\Sigma^{+}$ dominance takes place at relatively lower density $\rho
_{N} \approx 0.08 \fm ^{-3} \approx 0.5 \rho _{0}$.  This means that
the peak position of the $\LamStar$ spectrum in $K^{-} p \to (\pi
\Sigma)^{0}$ should be at an energy closed to the $\KbarN$ threshold
rather than at $1405 \mev$, because lower densities cannot prove the
energy far from the threshold.  In fact, we have an experimental
indication of the ratio increase and the inversion of the dominance
channel.  Namely, while the ratio $R_{+-}$ is $0.42$ for kaonic
hydrogen~\cite{Nowak:1978au, Tovee:1971ga}, which constrains the ratio
at the zero density, it becomes $0.85$ for kaonic deuterium, $1.8 \pm
0.5$ for kaonic $\HeF$~\cite{Katz:1970ng}, and $1.2$--$1.5$ for
$p$-shell nuclei~\cite{Agnello:2011iq}.  Therefore, experimental
results on $R_{+-}$ for various kaonic atoms could be explained by the
nature of the $\LamStar$ resonance.  More qualitative and quantitative
discussions on $K^{-}$ absorption in light kaonic atoms will be given
in Ref.~\cite{Sekihara:next}.

\subsection{Nonmesonic absorption}
\label{sec:two-body}

Next we show the results of the nonmesonic absorption potential of
$K^{-}$ calculated with the one-meson exchange model.  In the
nonmesonic absorption, the $\LamStar$ contribution appears in the
$K^{-}pp \to (Y N)^{+}$ and $K^{-}pn \to (Y N)^{0}$ processes whereas
the $K^{-}nn \to \Sigma ^{-} n$ process does not have the $\LamStar$
contributions within the one-meson exchange picture.  We also note
that we expect that the nonmesonic absorption potential would be
proportional to $\rho _{N}^{2}$, if there is no energy nor density
dependence in the $\KbarN$ amplitude.

The result of the nonmesonic absorption potential is shown in
Fig.~\ref{fig:abs-two}.  From the figure, we find that the absorption
potential has large contributions from the $K^{-}pp \to (YN)^{+}$ and
$K^{-}pn \to (YN)^{0}$ processes, while the $K^{-}nn\to \Sigma ^{-}n$
process gives tiny contribution.  Bearing in mind that $K^{-}$
absorption with a proton induces the $\LamStar$ resonance, we see that
these large contributions stem from the $\LamStar$ resonance and the
$\LamStar$ doorway process is realized also in the nonmesonic
absorption.

\begin{figure}[!Ht]
  \centering
    \Psfig{8.6cm}{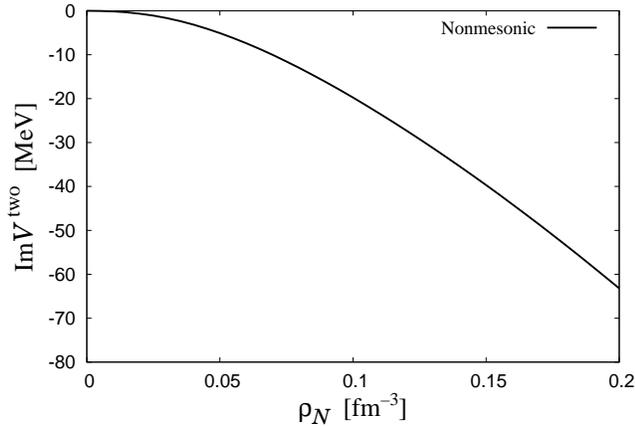}
    \caption{Total sum of nonmesonic absorption potential ($\text{Im}
      V^{\text{two}}$) for $K^{-}$ at rest in nuclear matter as a
      function of nuclear density.  }
  \label{fig:abs-two-total}
\end{figure}

\begin{figure}[!Ht]
  \centering
    \Psfig{8.6cm}{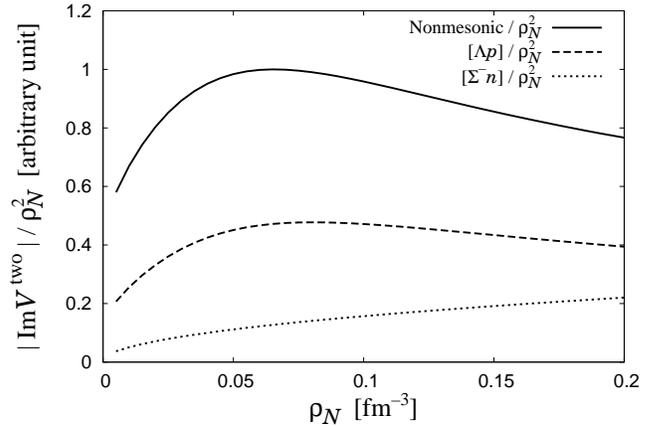}
    \caption{Absolute absorption potentials divided by $\rho _{N}^{2}$
      for total, $\Lambda p$, and $\Sigma ^{-} n$ contributions as
      functions of nuclear density in arbitrary unit.  Here potentials
      for the $\Lambda p$ and $\Sigma ^{-} n$ contributions are
      respectively multiplied by $5$ and $50$ relative to the total
      for comparison. }
  \label{fig:abs-two-rho}
\end{figure}

The total sum of the nonmesonic absorption potential is plotted in
Fig.~\ref{fig:abs-two-total} as a function of nuclear density.  The
total value of the nonmesonic absorption width ($=-2 \text{Im}
V^{\text{two}}$) amounts to about $100 \mev$ at the saturation density
$\rho _{0}=0.17 \fm ^{-3}$, although this value will be suppressed, as
in the mesonic absorption case, when the in-medium $\KbarN$ scattering
amplitude is employed.  The total nonmesonic absorption potential has
non-$\rho _{N}^{2}$ dependence, especially decreasing almost linearly
at high densities, due to the existence of the $\LamStar$ as doorway.
Then, in a similar manner to the mesonic absorption case, we can
extract the $\LamStar$ structure by evaluating the absolute nonmesonic
potentials divided by $\rho _{N}^{2}$, which contains information of
the squared amplitude $|{\cal T}|^{2}$ for the $K^{-}N\to MB$
transitions.  The result is plotted in Fig.~\ref{fig:abs-two-rho} for
the total and the $\Lambda p$ and $\Sigma ^{-} n$ nonmesonic channels.
From the figure, while no structure appears in the $\Sigma ^{-} n$
channel because of the absence of the $\LamStar$ contributions, the
total and $\Lambda p$ contributions show the peak structure around
$\rho _{N} \approx 0.06 \fm ^{-3} \approx 0.4 \rho _{0}$, which means
that the $\LamStar$ doorway is most prosperous at these densities
corresponding to the energy $1420 \mev$.  The peak position in
Fig.~\ref{fig:abs-two-rho} is consistent with the case of the mesonic
absorption potential.

\begin{figure}[!t]
  \centering
    \Psfig{8.6cm}{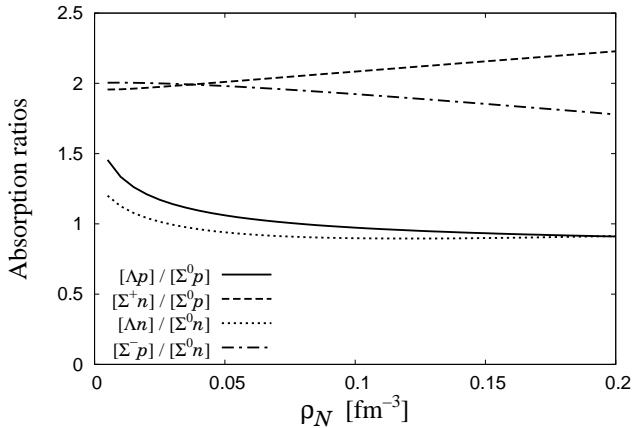}
    \caption{Ratios of nonmesonic absorption potentials for $[\Lambda
      p] / [\Sigma ^{0} p]$, $[\Sigma ^{+} n] / [\Sigma ^{0} p]$,
      $[\Lambda n] / [\Sigma ^{0} n]$, and $[\Sigma ^{-} p] / [\Sigma
      ^{0} n]$ as functions of nuclear density.  }
  \label{fig:ratio-two}
\end{figure}

The dominance of the $\LamStar$ contribution in the nonmesonic $K^{-}$
absorption can also be seen in the absorption ratios $[\Lambda p] /
[\Sigma ^{0} p]$, $[\Lambda n]/ [\Sigma ^{0} n]$, $[\Sigma
^{+}n]/[\Sigma ^{0}p]$, and $[\Sigma ^{-}p]/[\Sigma ^{0}n]$.  The
numerical results of our calculation for the $K^{-}$ absorption are
plotted as functions of nuclear density in
Fig.~\ref{fig:ratio-two}. As one can see, the absorption ratios
$[\Lambda p] / [\Sigma ^{0} p]$ and $[\Lambda n] / [\Sigma ^{0} n]$ in
our calculation show around unity almost independently of the density.
Bearing in mind that the previous study~\cite{Sekihara:2009yk} on the
$\LamStar N \to YN$ nonmesonic transition suggests the ratio of the
$\LamStar$ nonmesonic decays $[\Lambda N] / [\Sigma ^{0} N]$ to be
around $1.2$, one can see that the present results for $[\Lambda p] /
[\Sigma ^{0} p]$ and $[\Lambda n] / [\Sigma ^{0} n]$ are attributed to
the $\LamStar$ dominance in $K^-$ nonmesonic absorption.  Furthermore,
the $K^{-}$ absorption ratio $[\Sigma ^{+} n] / [\Sigma ^{0} p]$ and
$[\Sigma ^{-} p] / [\Sigma ^{0} n]$ are around two in these densities
in our calculation. This also suggests the $\LamStar$ dominance,
because if the initial $K^{-} p$ system is dominated by the $I=0$
component these ratios should be exactly two according to the isospin
symmetry.  Therefore, our result of the absorption ratios shows that
indeed the $\LamStar$ doorway process dominantly contributes to the
nonmesonic absorption of $K^{-}$ at rest in nuclear matter.

\subsection{Fractions of mesonic and nonmesonic absorptions}
\label{sec:branch}

\begin{figure}[!Ht]
  \centering
  \Psfig{8.6cm}{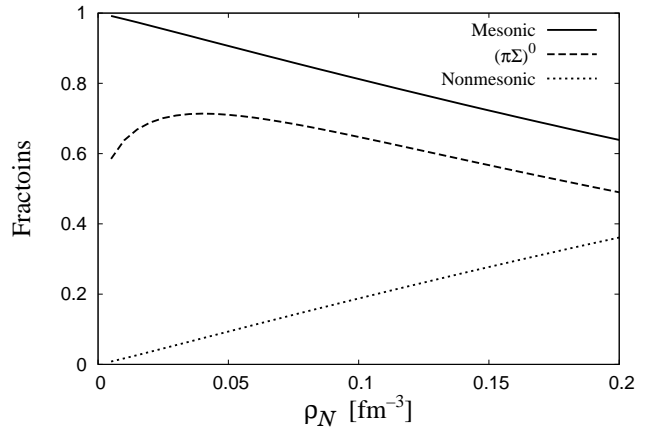}
  \caption{\label{fig:Branching_ratio} Fractions of mesonic, sum of
    $(\pi \Sigma )^{0}$, and nonmesonic absorption to total
    absorption. }
\end{figure}

Now it is interesting to compare the magnitude of the mesonic and
nonmesonic absorptions in our approach.  In order to see this, we show
in Fig.~\ref{fig:Branching_ratio} the fractions of the mesonic and
nonmesonic absorptions to the total, together with the fraction of the
sum of $(\pi \Sigma )^{0}$ states.  Here we note that although the
absorption potentials would be suppressed by the in-medium $\KbarN$
scattering amplitude, as discussed in preceding sections, the
fractions of the mesonic and nonmesonic absorptions to the total would
not largely change as long as the $\LamStar$ dominance would be valid.
As one can see from Fig.~\ref{fig:Branching_ratio}, the fraction of
the mesonic (nonmesonic) absorption almost linearly goes down (up)
from unity (zero) as the nuclear density increases.  The reason for
decrease (increase) of the fraction of the mesonic (nonmesonic)
absorption is that the nonmesonic reaction can more largely contribute
to the absorption at higher densities.  The almost linear dependence
of the fractions on density is a nontrivial result of the $\LamStar$
properties in the $\LamStar$ doorway process.  We also note that the
absorption to the $(\pi \Sigma )^{0}$ channels gives more than half of
the total absorption process.

\begin{figure}[!Ht]
  \centering
  \Psfig{8.6cm}{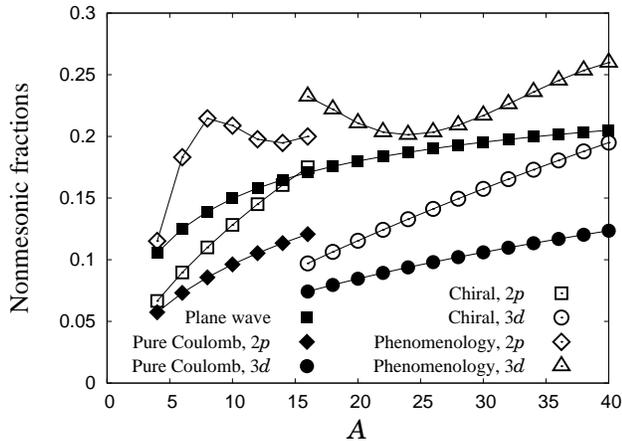}
  \caption{\label{fig:Branching_ratio_WS} Fraction of the nonmesonic
    absorption to the total absorption~\eqref{eq:pot_WS} for realistic
    nuclei with the Woods-Saxon densities.  Here we consider several
    wave functions for kaon: $2p$ and $3d$ of the wave function in the
    pure Coulomb potential and in the phenomenological and chiral
    unitary potentials in Ref.~\cite{Yamagata:2006sm}, and the plane
    wave.  }
\end{figure}

\begin{figure*}[!t]
  \centering
  \begin{tabular*}{\textwidth}{@{\extracolsep{\fill}}cc}
    \Psfig{8.6cm}{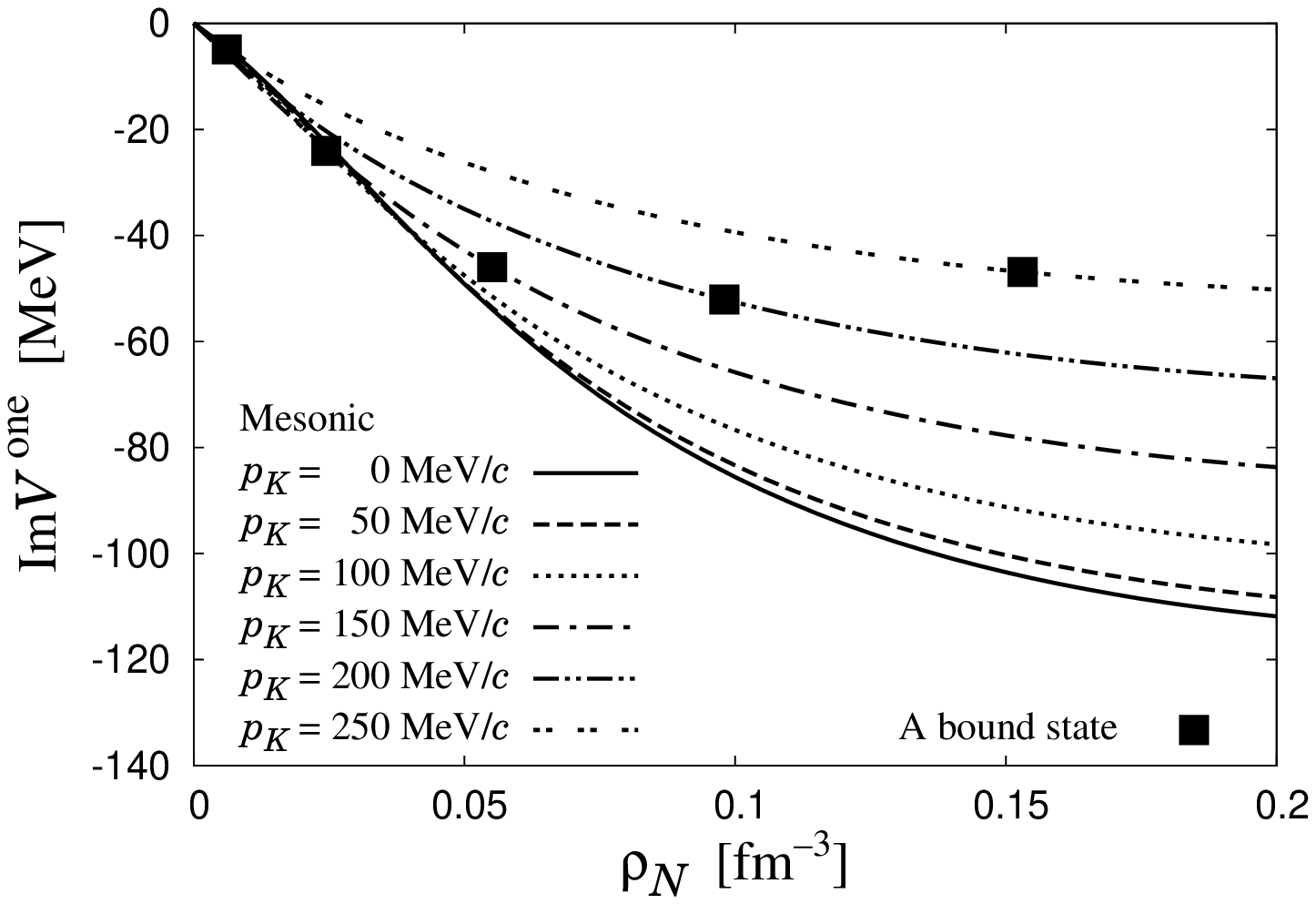} & 
    \Psfig{8.6cm}{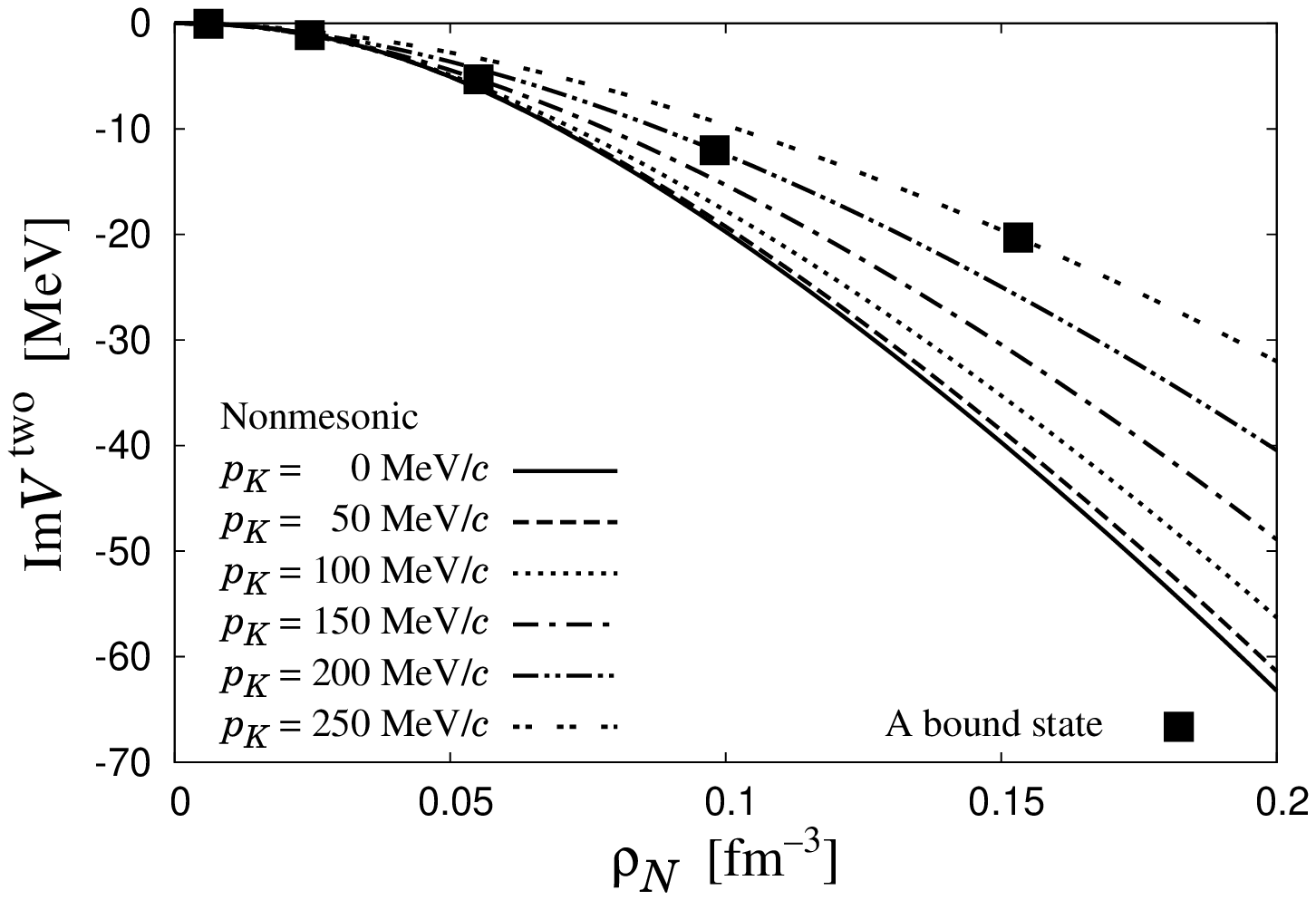} \\
    (a) & (b) 
  \end{tabular*}
  \caption{Mesonic (a) and nonmesonic (b) absorption pontentials with
    finite kaon momenta $p_{K^{-}}$.  A bound state case given in
    Eqs.~\eqref{eq:eigenvalue} and \eqref{eq:ReV} is indicated by the
    filled squares. }
  \label{fig:abs-pk} 
\end{figure*}%

Beside this, we emphasize that the mesonic and nonmesonic absorption
fractions are respectively about $70 \%$ and $30 \%$ at the saturation
density $\rho _{0} = 0.17 \fm ^{-3}$ of nuclear matter. This fraction
is close to the empirical value for kaonic atoms with nuclei heavier
than $\HeF$ (about $80 \%$ and $20 \%$,
respectively~\cite{Friedman:2007zza}).  From
Fig.~\ref{fig:Branching_ratio}, the nonmesonic fraction of $20 \%$
corresponds to $\rho _{N} = 0.1 \fm ^{-3} \approx 0.6 \rho _{0}$ in
our calculation of $K^{-}$ absorption at rest.

The absorption width for $K^{-}$ bound in finite nuclei is obtained as
the imaginary part of the eigenenergy of $K^-$. To obtain the
eigenenergy one solves equation of motion for the $K^-$-nucleus system
with the optical potential for $K^-$.  Here let us estimate the
nonmesonic absorption fraction for finite nuclei in an approximated
way based on a perturbation theory by calculating an overlap of the
absorption potential and a wave function for the bound $K^-$. To
evaluate the wave function we need both the real and imaginary parts
of the optical potential. Nevertheless, the calculation of the real
part of the optical potential is out of the scope of this work, so
that we take several examples for the $K^-$ wave function. It is worth
noting here that, to obtain the atomic wave funtion, one needs to
understand the energy spectrum of the $K^-$ nuclear states, since the
wave functions of the atomic and nuclear states should be orthogonal
if nuclear bound states exist, and the orthogonality condition is
significant for the behavior of the wave function in the region of the
nucleus size~\cite{Yamagata:2005ic}, where the absorption takes
place. In addition, owing to the orthogonality the wave functions of
the atomic states have nodes in the region of the nucleus, and this
implies that $K^-$ even in atomic states may have a large momentum
inside the nucleus, as suggested in Refs.~\cite{Cieply:2011yz,
  Cieply:2011fy}. It is also known that the effective density where
the absorption takes place mainly is strongly dependent on the strong
interaction between $K^-$ and nucleus~\cite{Yamagata:2006sv,
  Yamagata-Sekihara:HYP}.

From the nuclear density distribution, we assume the Woods-Saxon form
\begin{equation}
\rho _{\rm WS} ( r ) 
\equiv \frac{\bar{\rho}}{1 + \exp [ ( r - R ) / a]} ,
\end{equation}
where we take the nuclear radius $R=1.18 A^{1/3} - 0.48 \fm$ and the
diffuseness $a=0.5 \fm$, which reproduce empirical density
distributions of nuclei, and the normalization $\bar{\rho}$ is fixed
so as to reproduce the atomic number $A$,
\begin{equation}
A = \int d^{3} r \rho _{\rm WS} ( r ) . 
\end{equation}

Applying the local density approximation, we evaluate the absorption
width as
\begin{equation}
  \frac \Gamma 2= {\cal N} \int d^{3} r \, |\psi (r)|^{2}
  \text{Im} V ( \rho _{\rm WS} ( r ) ) ,
\label{eq:pot_WS}
\end{equation}
where $\psi (r)$ is the $K^{-}$ wave function.  Here we consider
several wave functions $\psi (r)$ in the $2p$ and $3d$ states, which
are obtained by the pure Coulombic potential, the phenomenological
potential and chiral unitary model. The latter two potentials are
discussed in Ref.~\cite{Yamagata:2006sm}. We also consider a plane
wave with zero momentum, which could be the case of in-flight kaons
with very low momentum, such $10 \mev /c$.

We show in Fig.~\ref{fig:Branching_ratio_WS} the result of the
nonmesonic absorption fraction to the total for nuclei of $A=4$--$40$
with assumption $Z=N$. As one can see, the fractions of the nonmesonic
absorption to the total absorption are marginally dependent on the
wave functions. For the detail discussion, one needs to evaluate the
wave functions in a more appropriate way using a realistic optical
potential including the momentum dependence.

\subsection{Absorption with finite kaon momenta and energies}
\label{sec:FiniteEP} 

Until the previous subsections we have considered the self-energy of
kaon at rest in nuclear matter with $p_{K^{-}}^{\mu}=(m_{\bar{K}}, \,
\bm{0})$.  In this subsection let us take into account the finite kaon
momenta $\bm{p}_{K^{-}} \neq \bm{0}$ and energies
$E_{K^{-}}<m_{\bar{K}}$.  These effect will be important to
investigate absorption of kaon into actual finite nuclei, because
attractive strong interaction will change the kaon momentum as well as
the energy from zero to finite values at the absorption point.
Especially, for the atomic states, kaons in the center of the nucleus
may have large momenta to compensate a large strong attraction by the
kaon kinetic energy for small atomic binding energy as suggested in
Ref.~\cite{Cieply:2011yz, Cieply:2011fy}.

One important influence of the finite kaon momenta and energies is the
downawrd shift of the $K^{-} N$ two-body energy $W$~\eqref{eq:W-one}
due to the off-shellness of kaon. Actually for the kaon
energy-momentum $p_{K^{-}}^{\mu}=(E_{K^{-}}, \, \bm{p}_{K^{-}})$ the
two-body energy $W$ becomes, after averaging the angular dependence,
\begin{equation}
W = \sqrt{(E_{1} + E_{K^{-}} )^{2} 
- p_{1}^{2} - p_{K^{-}}^{2} } , 
\end{equation}
which is obviously smaller than $W$ with $E_{K^{-}}=m_{\bar{K}}$ and
$\bm{p}_{K^{-}} = \bm{0}$.  This fact indicates that the nuclear
density which hits $\LamStar$ will become lower according to the
values of $p_{K^{-}}^{2}$ and $E_{K^{-}}$.  Here we will see how this
two-body energy shift affects the absorption scenarios with finite
kaon momenta and energies.

\begin{figure*}[!t]
  \centering
  \begin{tabular*}{\textwidth}{@{\extracolsep{\fill}}cc}
    \Psfig{8.6cm}{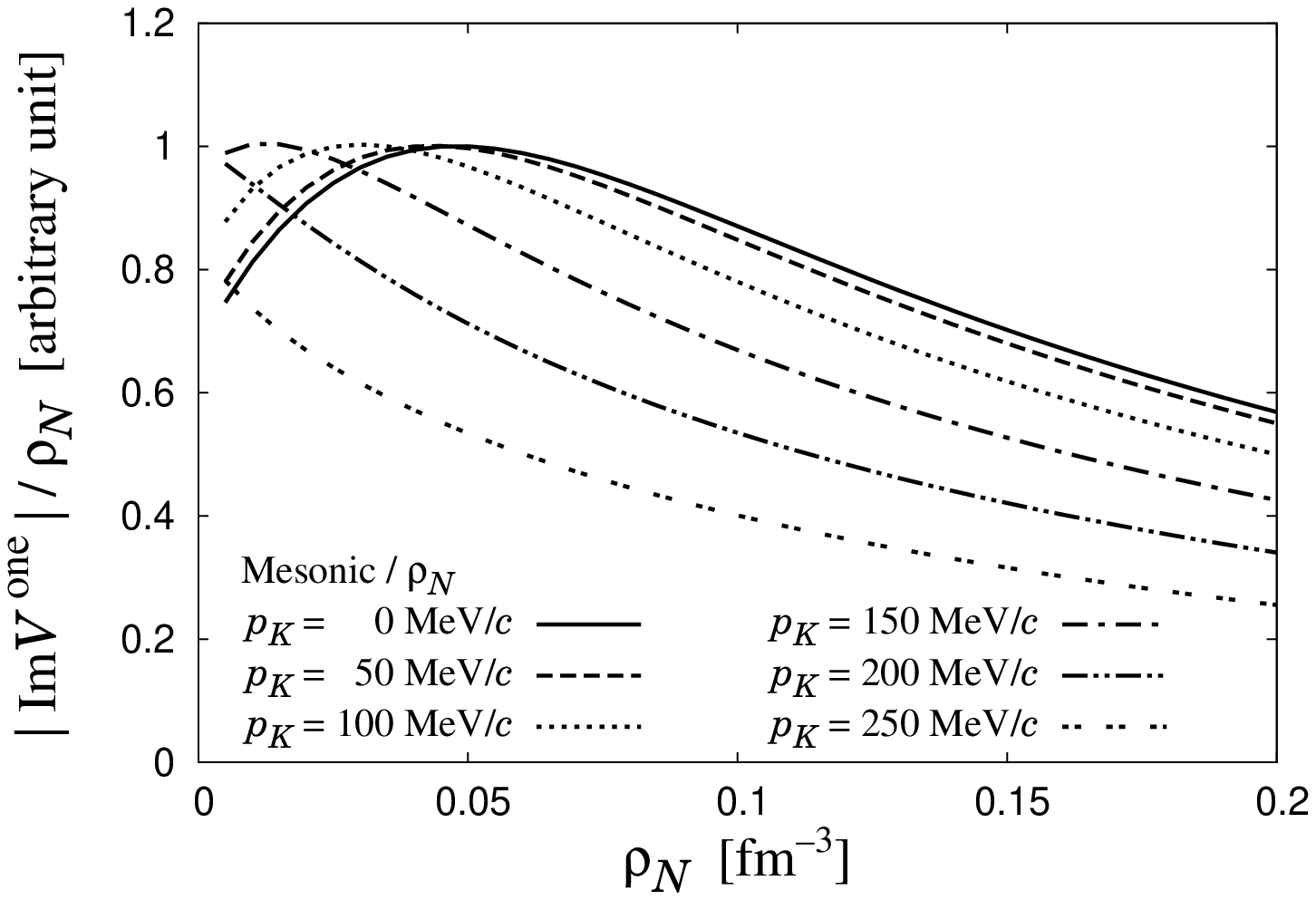} & 
    \Psfig{8.6cm}{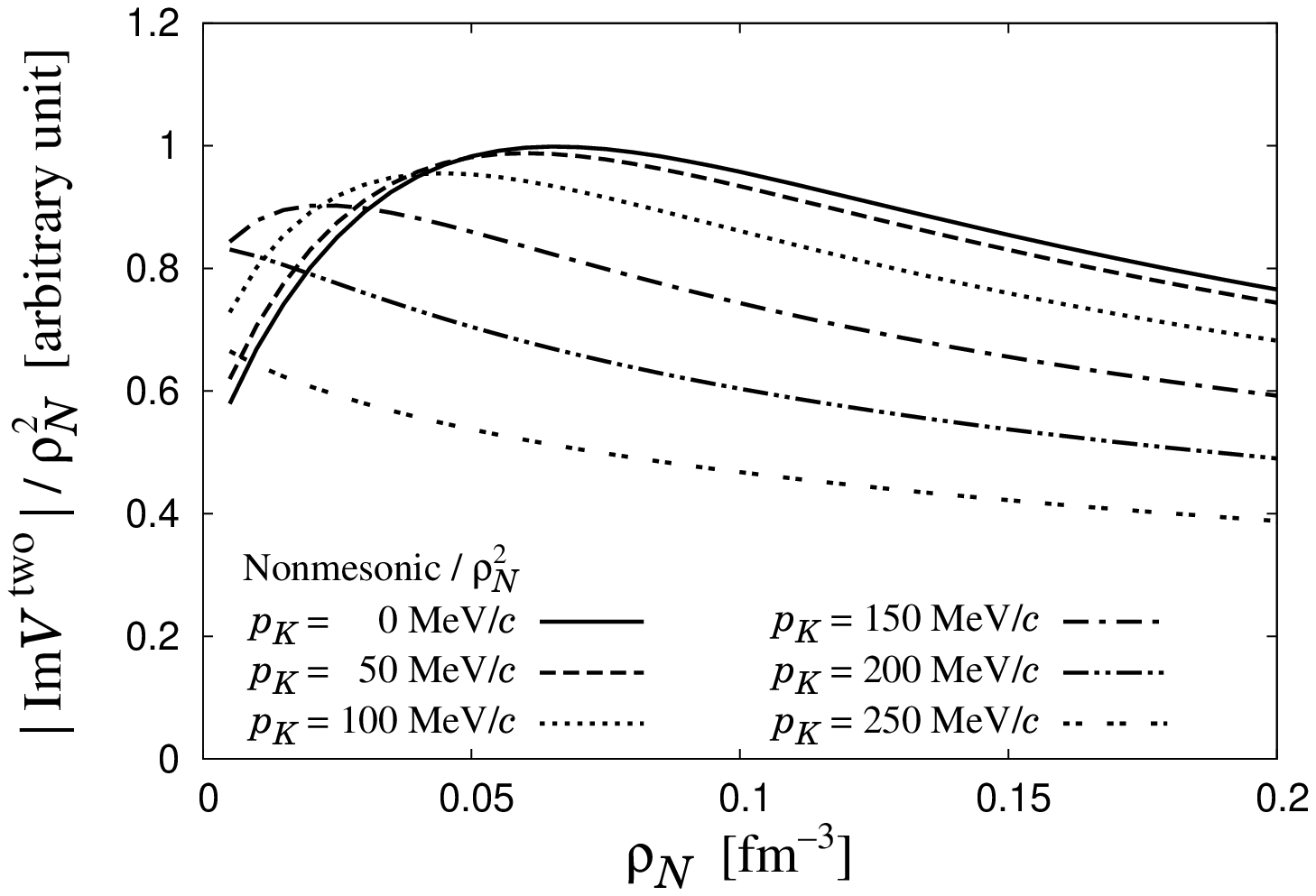} \\
    (a) & (b)
  \end{tabular*}
  \caption{Absolute absorption potentials with finite kaon momenta
      $p_{K^{-}}$ divided by $\rho _{N}$ for mesonic case (a) and
      divided by $\rho _{N}^{2}$ for nonmesonic case (b) in arbitrary
      unit. }
  \label{fig:abs-rho-pk} 
\end{figure*}%

\begin{figure}[!Ht]
  \centering
  \Psfig{8.6cm}{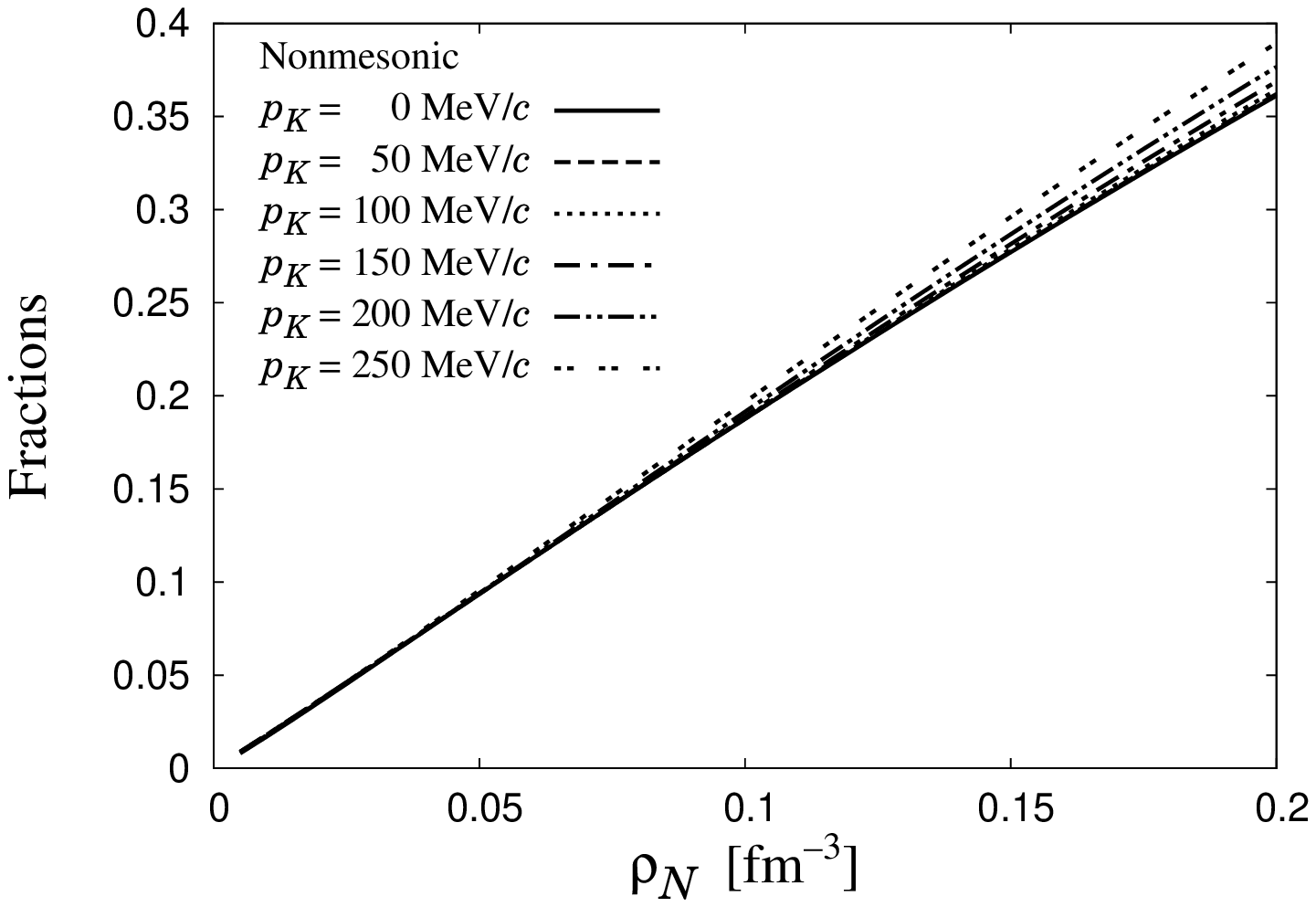}
  \caption{Fraction of the nonmesonic absorption to the total
    absorption with finite kaon momenta $p_{K^{-}}$. }
  \label{fig:ratio-pk}
\end{figure}

Firstly we consider the finite kaon momenta with
$p_{K^{-}}^{\mu}=(m_{\bar{K}}, \, \bm{p}_{K^{-}})$.  Here we take the
approximation that we average the angular dependence appearing in the
$K^{-}NN$ three-body energy $E_{\rm tot}$ in the nonmesonic absorption
as well as the $K^{-} N$ two-body energy $W$ so that one drops the
angular dependence of $\bm{p}_{K^{-}} \cdot \bm{p}_{1}$ and
$\bm{p}_{K^{-}} \cdot \bm{p}_{2}$ with initial nucleon momenta
$\bm{p}_{1}$ and $\bm{p}_{2}$.  The mesonic and nonmesonic absorption
potentials with finite kaon momenta are shown in Fig.~\ref{fig:abs-pk}
from $p_{K^{-}}=0 \mev /c$, which is the same case as the previous
subsections, to $p_{K^{-}}=250 \mev /c$.  As one can see, the
absorption potentials become weaker as the kaon momentum increases in
both mesonic and nonmesonic cases.  While with the small kaon momenta
$p_{K^{-}} \lesssim 100 \mev /c$ the absorption potentials are
suppressed only slightly, the potentials with $p_{K^{-}} \gtrsim 200
\mev /c$ become about halft of the potential with kaon at rest at the
nuclear saturation density.  

For the bound state, since it is an eigenstate, the momentum and
position of the kaon are correlated. If one takes the local density
approximation, which connects position and density, the density and
momentum can be also correlated. Therefore, each bound state may have
one line for the absorption strength against the density. As an
example, we put squares in Fig.~\ref{fig:abs-pk} for a bound kaon
atomic state calculated with a density-momentum relation,
\begin{equation}
  \frac{p_{K^{-}} (\rho _{N} )^{2}}{2 m_{\bar{K}}} 
  + \text{Re} V ( \rho _{N} ) = - \text{(atomic binding energy)} \approx 0 .
  \label{eq:eigenvalue}
\end{equation}
where we take a potential proportional to the nuclear density with a
typical potential strength from the chiral unitary aproach,
\begin{equation}
  \text{Re} V( \rho _{N} ) = -70 \mev \times \frac{\rho _{N}}{\rho_0} . 
  \label{eq:ReV}
\end{equation}
The squares show that the growth of the absorption potentials become
decrease as the nuclear density gets large compared to the case of
kaon at rest due to the increase of $p_{K^{-}}$ as a function of the
nuclear density.  The squares will move upward (downward) in the
figure as the potential strength becomes strong (weak). 

The suppression of the absorption potential is caused by the two
reasons.  One is, as we have already mentioned, the downward shift of
the two-body energy $W$ due to the finite $p_{K^{-}}$ makes the
nuclear density which hits the $\LamStar$ resonance lower, and hence
the $\LamStar$ doorway becomes weak compared to the case of kaon at
rest at the saturation density.  Indeed, we can estimate the density
at which the $\LamStar$ contribution is large by calculating
$|\text{Im} V^{\text{one}}| / \rho _{N}$ and $|\text{Im}
V^{\text{two}}| / \rho _{N}^{2}$, and the results with finite kaon
momenta is plotted in Fig.~\ref{fig:abs-rho-pk}.  From the figure, one
can see the peak position shifts downward density as the kaon momentum
increases, which is consistent with the expectation from behavior of
$W$, and the peak disappears at $p_{K^{-}} \sim 200 \mev /c$ because
in such kaon momenta the two-body energy $W$ is smaller than the
$\LamStar$ peak position even in the low density limit, $\rho _{N}\to
0$.  The other reason for the suppression of the absorption potential
is that the downward shift of $W$ makes the phase space for the
on-shell $\pi Y$ mesonic channels and $Y N$ nonmesonic channels small
and hence suppresses the reaction rate for the absorption, $\gamma
_{\pi Y, \, Y N}$.  We have checked that these two factors suppresses
the absorption potential with similar strength.  We note that the
phase-space suppression is especially crusial to the mesonic
absorption because the $W$ is closer to the $\pi Y$ threshold in the
mesonic case than the $E_{\rm tot}$ to the $YN$ threshold in the
nonmesonic case.

The fraction of nonmesonic absorption to the total absorption with
finite kaon momenta is plotted in Fig.~\ref{fig:ratio-pk}.  The figure
indicates that, although the absolute absorption potentials are
suppressed due to the finite kaon momenta both in the mesonic and
nonmesonic cases, the fraction only slightly changes because of the
cancellation of the suppressions.  This means that the results for the
nonmesonic fraction obtained in the previous subsection is not so
sensitive to the kaon momentum.

\begin{figure*}[!Ht]
  \centering
  \begin{tabular*}{\textwidth}{@{\extracolsep{\fill}}cc}
    \Psfig{8.6cm}{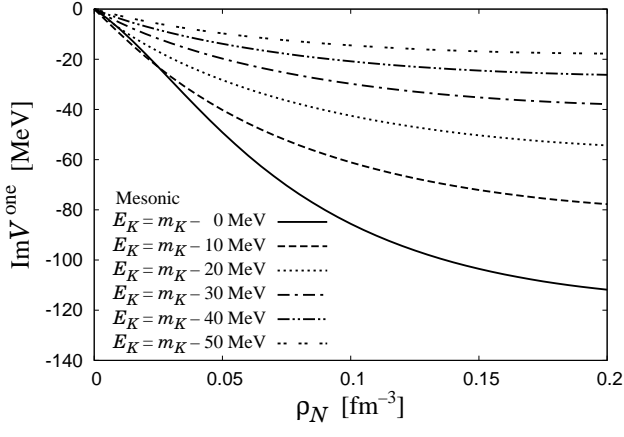} & 
    \Psfig{8.6cm}{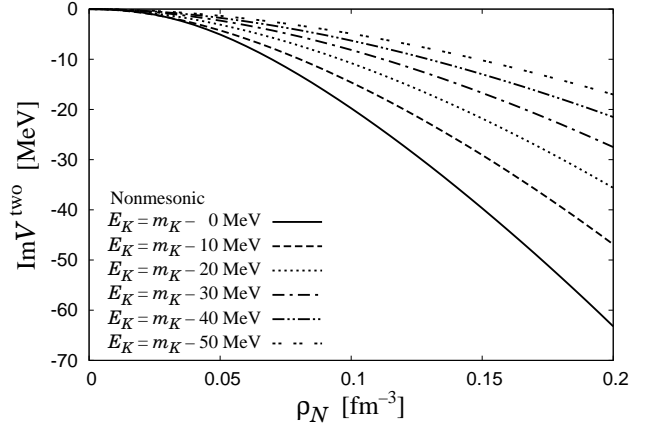} \\
    (a) & (b)
  \end{tabular*}
  \caption{Mesonic (a) and nonmesonic (b) absorption pontentials
      with kaon energy $E_{K^{-}}<m_{\bar{K}}$. } 
  \label{fig:abs-Bk} 
\end{figure*}%

\begin{figure}[!Ht]
  \centering
  \Psfig{8.6cm}{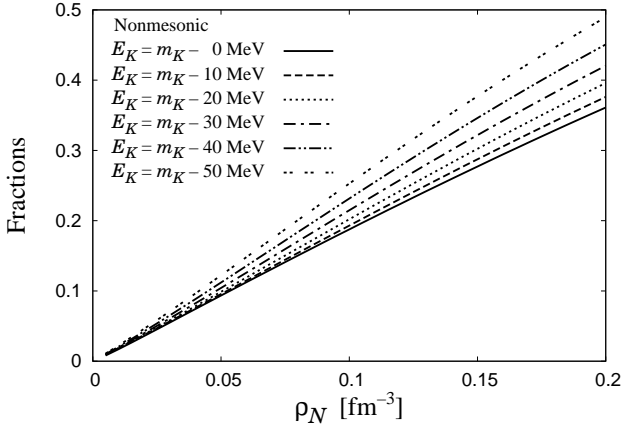}
  \caption{Fraction of the nonmesonic absorption to the total
    absorption with kaon energy $E_{K^{-}}<m_{\bar{K}}$. }
  \label{fig:ratio-Bk}
\end{figure}

Next let us take into account the kaon energy.  Here we assume the
kaon is at rest with energy $E_{K^{-}}<m_{\bar{K}}$,
$p_{K^{-}}^{\mu}=(E_{K^{-}}, \, \bm{0})$.  The mesonic and nonmesonic
absorption potentials with the finite kaon energies are plotted in
Fig.~\ref{fig:abs-Bk} up to $m_{\bar{K}}-50 \mev$.  From the figure,
one can see the suppression for the absorption potentials in a similar
manner to the finite kaon momentum case.  For the finite kaon energy,
the absorption potentials are largely suppressed even at
$E_{K^{-}}=m_{\bar{K}}-10 \mev$, which reflects that the energy shift
due to that energy is large enough to suppress the $\LamStar$ doorway
contribution and the phase space for the decay channel.  In both
mesonic and nonmesonic cases, the absorption potentails at the
saturation density becomes half for the kaon energy $E_{K^{-}} \sim
m_{\bar{K}}-20 \mev$ compared to the potential for kaon with
$E_{K^{-}}=m_{\bar{K}}$.  The fraction of nonmesonic absorption to the
total absorption with finite kaon energies is plotted in
Fig.~\ref{fig:ratio-Bk}.  The nonmesonic fraction increases as the
energy decreases, and at the saturation density the fraction becomes
$\sim 0.4$ with $E_{K^{-}}=m_{\bar{K}}-50 \mev$ while it is $\sim 0.3$
for kaon with $E_{K^{-}}=m_{\bar{K}}$.

Finally we summarize our results for the $K^{-}$ absorption potential
with the $s$-wave $\KbarN \to MB$ transition amplitude.  We have seen
that $K^{-}$ absorption at rest is dominated by the $\LamStar$ doorway
process, where the transitions of the initial state $K^{-}N$ to $MB$
take place mainly through the $\LamStar$ resonance.  From the behavior
of the absorption potential the $\LamStar$ contributes mostly at the
nuclear density $\rho _{N} \approx 0.05$--$0.06 \fm ^{-3}$.  We have
found that increase of the ratio $[\pi ^{-} \Sigma ^{+}] / [\pi ^{+}
\Sigma ^{-}]$ in experiments of heavier kaonic atoms can be explained
as the interference with the nonresonant $I=1$ background with respect
to the $\LamStar$ contributions in the $\KbarN$ subthreshold region.
Due to the dominance of the $\LamStar$ doorway process, the nonmesonic
absorption ratios $[\Lambda p]/ [\Sigma ^{0} p]$ and $[\Lambda n]/
[\Sigma ^{0} n]$ are about unity while $[\Sigma ^{+}n]/[\Sigma ^{0}p]$
and $[\Sigma ^{-}p]/[\Sigma ^{0}n]$ are about two.  In addition, our
approach gives that the mesonic and nonmesonic absorption fractions
are respectively about $70 \%$ and $30 \%$ at the saturation density.
Estimating the surface effect for finite nuclei with some examples of
the $K^-$ wave function, we have found that the fraction of the
nonmesonic absorption will be about $10$--$20 \%$.  The details are
dependent on the atomic wave function, and thus more realistic
evaluation is necessary with real part of the optical potential.
Taking into account the kaon momenta and the kaon energies, the
absorption potentials become weaker due to the downward shift of the
$K^{-} N$ two-body energy.  However, even in such a case the fraction
of the nonmesonic absorption does not drastically change because of
the cancellation of the suppressions of potentials.

\section{$\bm{\SigTEF}$ contributions}
\label{sec:Sigma1385}

Next we examine the $\SigTEF$ ($\SigStar$) contribution to $K^{-}$
absorption.  The hyperon resonance $\SigStar$ exists below the
$\KbarN$ threshold and couples to $\KbarN$ and $\pi Y$ channels in $p$
wave.  Since the scattering amplitude with $p$-wave coupling is
proportional to the momentum transfer, we expect that the $\SigStar$
contribution to the $K^{-}$ absorption with small kaon momenta is
small compared to the $\LamStar$ contribution, which couples to
$\KbarN$ and $\pi \Sigma$ channels in $s$ wave.  Here we also discuss
the $\SigStar$ nonmesonic decay in nuclear matter in the similar
manner to the $\LamStar$ resonance developed in
Ref.~\cite{Sekihara:2009yk}.  Because we are interested in the $K^{-}$
absorption in nuclear matter, we take into account $\Sigma ^{\ast 0}$
and $\Sigma ^{\ast -}$ contributions, while $\Sigma ^{\ast +}$ is not
considered in this study since it is not directly produced in the
$K^{-} N$ initial state.  Throughout this study we neglect in-medium
modifications on $\SigStar$.

\subsection{$\bm{\SigTEF}$-induced nonmesonic decay}

Before going to the $K^{-}$ absorption, we discuss the nonmesonic
decay process of $\SigStar$ in nuclear matter by considering the
$\SigStar N \to YN$ transition in the nuclear medium.  This enables us
to investigate the nonmesonic decay pattern for the $\SigStar$
dominance, and is a supplemental study with respect to the
$\LamStar$-induced nonmesonic decay discussed in
Ref.~\cite{Sekihara:2009yk}.  For this purpose, we calculate the
$\SigStar N \to Y N$ process (%
$\Sigma^{\ast 0} p \to \Lambda p$, $\Sigma ^{0} p$, and $\Sigma ^{+}
n$, %
$\Sigma ^{\ast 0}n \to \Lambda n$, $\Sigma ^{0} n$, and $\Sigma ^{-}
p$, %
$\Sigma ^{\ast -}p \to \Lambda n$, $\Sigma ^{0} n$, and $\Sigma ^{-}
p$, %
and %
$\Sigma ^{\ast -} n \to \Sigma ^{-} n$) %
in uniform nuclear matter with a one-meson exchange approach, as done
in Ref.~\cite{Sekihara:2009yk}.  Here we note that we have two cases
of initial states, $\Sigma ^{\ast 0} n$ and $\Sigma ^{\ast -} p$, for
the $(YN)^{0}$ final states.

\begin{figure}[!t]
  \bc
    \includegraphics[scale=0.17]{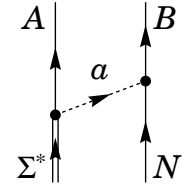}
    \caption{Feynman diagram for the $\SigStar N \to YN$ process.  The
      propagating particles $a$, $A$, and $B$ are listed in
      Table~\ref{tab:Feynman-Sigma}. }
  \label{fig:Feynman-Sigma}
  \ec
\end{figure}

\begin{table}[t]
  \caption{\label{tab:Feynman-Sigma} 
    Possible channels for Eq.~\eqref{eq:Bchannels}.  Here 
    $\SigStar$ and $N$ are
    the hyperon resonance and nucleon in the initial state, while $A$ and
    $B$ are the baryons in the final state. $a$ denotes the exchange
    meson.  $\xi$ is the relative sign of the amplitude coming from the
    exchange of the final state baryons. $C$ is the Clebsch-Gordan
    coefficients for the $a A \SigStar$ coupling, and $\alpha$ and $\beta$
    are the Clebsch-Gordan coefficients for the $BBM$ coupling. }
  \begin{ruledtabular}
    \begin{tabular}{ccccccccc} 
      $\SigStar$ & $N$ & $a$ & $A$ & $B$ & $\xi$ & $C_{aA}$ 
      & $\alpha$ & $\beta$
      \\ \hline
      $\Sigma ^{\ast 0}$ & $p$ & 
      \rule[0pt]{0pt}{9pt} $K^{-}$ & $p$ & $\Lambda$ 
      & $+$ & $-\sqrt{1/12}$ & $-2/\sqrt{3}$ & $1/\sqrt{3}$ \\
      & & $\pi ^{0}$ & $\Lambda$ & $p$ 
      & $-$ & $1/2$ & $1$ & $0$ \\ \hline 
      $\Sigma ^{\ast 0}$ & $p$ & 
      \rule[0pt]{0pt}{9pt} $K^{-}$ & $p$ & $\Sigma ^{0}$ 
      & $+$ & $-\sqrt{1/12}$ & $0$ & $1$ \\
      & & $\eta$ & $\Sigma ^{0}$ & $p$ 
      & $-$ & $-1/2$ & $1/\sqrt{3}$ & $-2/\sqrt{3}$ \\ \hline 
      $\Sigma ^{\ast 0}$ & $p$ & 
      \rule[0pt]{0pt}{9pt} $\bar{K}^{0}$ & $n$ & $\Sigma ^{+}$ 
      & $+$ & $\sqrt{1/12}$ & $0$ & $\sqrt{2}$ \\
      & & $\pi ^{-}$ & $\Sigma ^{+}$ & $n$ 
      & $-$ & $\sqrt{1/12}$ & $\sqrt{2}$ & $0$ \\ 
      \hline \hline 
      $\Sigma ^{\ast 0}$ & $n$ & 
      \rule[0pt]{0pt}{9pt} $\bar{K}^{0}$ & $n$ & $\Lambda$ 
      & $+$ & $\sqrt{1/12}$ & $-2/\sqrt{3}$ & $1/\sqrt{3}$ \\
      & & $\pi ^{0}$ & $\Lambda$ & $n$ 
      & $-$ & $1/2$ & $-1$ & $0$ \\ \hline 
      $\Sigma ^{\ast 0}$ & $n$ & 
      \rule[0pt]{0pt}{9pt} $\bar{K}^{0}$ & $n$ & $\Sigma ^{0}$ 
      & $+$ & $\sqrt{1/12}$ & $0$ & $-1$ \\
      & & $\eta$ & $\Sigma ^{0}$ & $n$ 
      & $-$ & $-1/2$ & $1/\sqrt{3}$ & $-2/\sqrt{3}$ \\ \hline 
      $\Sigma ^{\ast 0}$ & $n$ & 
      \rule[0pt]{0pt}{9pt} $K^{-}$ & $p$ & $\Sigma ^{-}$ 
      & $+$ & $-\sqrt{1/12}$ & $0$ & $\sqrt{2}$ \\
      & & $\pi ^{+}$ & $\Sigma ^{-}$ & $p$ 
      & $-$ & $-\sqrt{1/12}$ & $\sqrt{2}$ & $0$ \\ 
      \hline \hline 
      $\Sigma ^{\ast -}$ & $p$ & 
      \rule[0pt]{0pt}{9pt} $K^{-}$ & $n$ & $\Lambda$ 
      & $+$ & $-\sqrt{1/6}$ & $-2/\sqrt{3}$ & $1/\sqrt{3}$ \\
      & & $\pi ^{-}$ & $\Lambda$ & $n$ 
      & $-$ & $1/2$ & $\sqrt{2}$ & $0$ \\ \hline 
      $\Sigma ^{\ast -}$ & $p$ & 
      \rule[0pt]{0pt}{9pt} $K^{-}$ & $n$ & $\Sigma ^{0}$ 
      & $+$ & $-\sqrt{1/6}$ & $0$ & $1$ \\
      & & $\pi ^{-}$ & $\Sigma ^{0}$ & $n$ 
      & $-$ & $-\sqrt{1/12}$ & $\sqrt{2}$ & $0$ \\ \hline 
      $\Sigma ^{\ast -}$ & $p$ & 
      \rule[0pt]{0pt}{9pt} $\pi ^{0}$ & $\Sigma ^{-}$ & $p$ 
      & $-$ & $\sqrt{1/12}$ & $1$ & $0$ \\
      & & $\eta$ & $\Sigma ^{-}$ & $p$ 
      & $-$ & $-1/2$ & $1/\sqrt{3}$ & $-2/\sqrt{3}$ \\
      \hline \hline 
      $\Sigma ^{\ast -}$ & $n$ & 
      \rule[0pt]{0pt}{9pt} $K^{-}$ & $n$ & $\Sigma ^{-}$ 
      & $+$ & $-\sqrt{1/6}$ & $0$ & $\sqrt{2}$ \\
      & & $\pi ^{0}$ & $\Sigma ^{-}$ & $n$ 
      & $-$ & $\sqrt{1/12}$ & $-1$ & $0$ \\ 
      & & $\eta$ & $\Sigma ^{-}$ & $n$ 
      & $-$ & $-1/2$ & $1/\sqrt{3}$ & $-2/\sqrt{3}$ \\ 
    \end{tabular}
  \end{ruledtabular}
\end{table}%

\begin{figure*}[!Ht]
  \centering
  \begin{tabular*}{\textwidth}{@{\extracolsep{\fill}}cc}
    \Psfig{8.6cm}{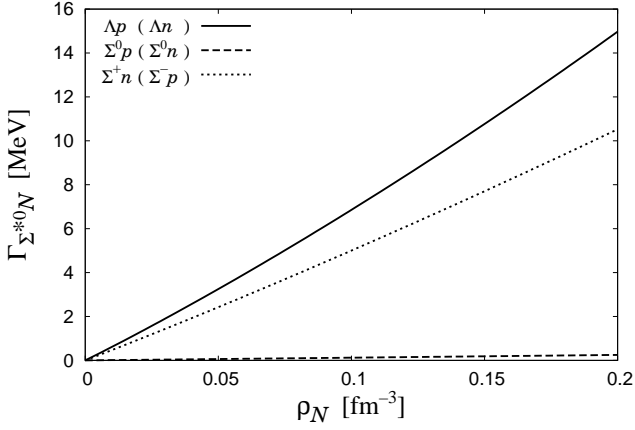} & 
    \Psfig{8.6cm}{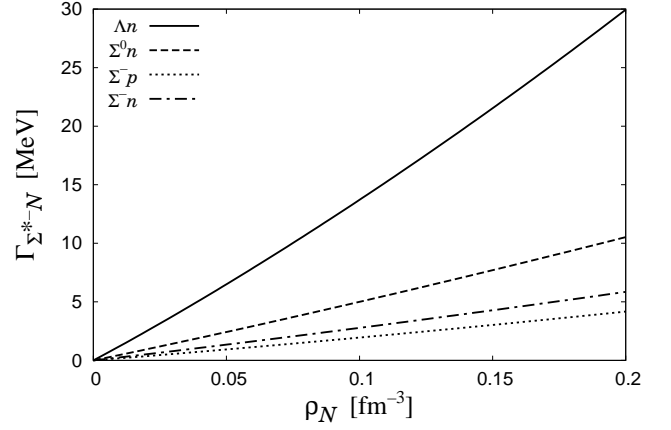} \\
    (a) & 
    (b)
  \end{tabular*}
  \caption{The $\SigStar$-induced nonmesonic decay width as a function
    of nuclear density.  (a) and (b) shows the $\Sigma ^{\ast 0}$ and
    $\Sigma ^{\ast -}$ contributions, respectively.  }
  \label{fig:width-sig}
\end{figure*}

In this study we use one-meson exchange model diagrammatically 
shown in Fig.~\ref{fig:Feynman-Sigma} with propagating particles 
listed in Table~\ref{tab:Feynman-Sigma}.  Along with the 
$\LamStar$-induced nonmesonic decay discussed in 
Ref.~\cite{Sekihara:2009yk}, we define the nonmesonic decay width 
of $\SigStar$ in nuclear matter through the $\SigStar N \to YN$ 
process, $\Gamma _{\SigStar N \to YN}$, as, 
\be
\Gamma _{\SigStar N \to YN} 
= \int _{0}^{k_{\text{F}}} \frac{d p_{1} \, p_{1}^{2}}{\pi ^{2}} 
\overline{\sum_{\lambda}} \sum _{\lambda ^{\prime}} 
\gamma _{\SigStar N \to YN} , 
\ee
\be
\gamma _{\SigStar N \to YN} \equiv 
\frac{p_{\text{cm}}^{\prime \prime} M_{Y} M_{N}}{4 \pi ^{2} E_{\text{tot}}} 
\int d \Omega _{N} 
\left | 
{\cal B} _{Y N} 
\right | ^{2} , 
\ee
where ${\cal B}_{YN}$ is the scattering amplitude for the 
$\SigStar N \to YN$ process written as, 
\begin{equation}
{\cal B}_{Y N} = 
\sum _{i} \xi _{i} {\cal B}_{i} 
(\Sigma _{i}^{\ast} N_{i} \xrightarrow{a_{i}} A_{i} B_{i})
\label{eq:Bchannels}
\end{equation}
for channel $i$ with
an amplitude for $\SigStar N \xrightarrow{a} AB$ process, 
\be
{\cal B} ( \SigStar N \xrightarrow{a} AB ) 
= i D_{aA} 
\times \tilde{\Pi} _{a}^{(p)} 
(q_{a}^{2}; \, \bm{q}_{a}, \, \bm{S}_{1}, \, \bm{\sigma}_{2}) 
\times \tilde{V}_{aNB} . 
\label{eq:Belement}
\ee
Here the symbol $\xi$ denotes relative sign of the amplitude coming
from the exchange of the final-state baryons, $a$ denotes the
propagating meson, and $D_{aA}$ is the $aA \SigStar$ coupling constant,
which we evaluate by first using the SU(6) quark model to relate the
$\pi N N$ coupling to the $\pi N \Delta$ one and then using the flavor
SU(3) symmetry to relate the $\pi N \Delta$ coupling to the $\pi Y
\SigStar$, $\eta Y \SigStar$, and $\bar{K} N \SigStar$ ones, as done
in Ref.~\cite{Oset:2000eg}, and as a result we obtain,
\be
D_{aA} = C_{aA} \frac{12}{5} \frac{D+F}{2 f} , 
\label{eq:Sig-coup}
\ee
with the SU(3) coefficient $C_{aA}$ listed in
Table~\ref{tab:Feynman-Sigma} and parameters $D+F=1.26$ and
$f=f_{\pi}=93.0 \mev$.  The propagator with $p$-wave short-range
correlation $\tilde{\Pi} _{a}^{(p)}$ is written as~\cite{Oset:1979bi},
\begin{align}
& \tilde{\Pi} _{a}^{(p)} (q^{2}; \, \bm{q}, \, \bm{S}, \, \bm{\sigma}) 
\nonumber \\ & 
= 
(\bm{q} \cdot \bm{S}) 
(\bm{q} \cdot \bm{\sigma}) 
\tilde{\Pi} _{a} (q^{2}) 
- 
\bm{S} 
\cdot 
\bm{\sigma} 
\frac{q_{\text{C}}^{2}}{3} 
\left ( \frac{\Lambda ^{2}}{\Lambda ^{2} - \tilde{q}^{2}} \right ) ^{2} 
\frac{1}{\tilde{q}^{2} - m_{a}^{2}} . 
\end{align}
We use here the same parameters $\Lambda = 1.0 \gev$ and
$q_{\text{C}}=780 \mev$ as for the $s$-wave short-range correlations.
For the $MBB$ coupling $\tilde{V}$ we use the same one 
as in the previous section, 
\begin{equation}
\tilde{V}_{a N B}
= \alpha _{a N B} \frac{D + F}{2 f} + \beta _{a N B} \frac{D -
  F}{2 f} .
\end{equation}
The vector $\bm{S}$ is the spin transition operator from spin $3/2$ to
$1/2$ having a relation,
\be
S^{i} S^{j \dagger} 
= \frac{2}{3} \delta ^{ij} - \frac{i}{3} \epsilon _{ijk} \sigma ^{k} . 
\ee
The subscript $1$ ($2$) for the operator $\bm{S}$ ($\bm{\sigma}$) in
Eq.~\eqref{eq:Belement} means that the operator is sandwiched by the
spinors for $\SigStar$ and $A$ ($N$ and $B$).  The $\SigStar$ mass 
is fixed as $1385 \mev$. 

The results of the nonmesonic decay width of $\Sigma ^{\ast 0}$ and
$\Sigma ^{\ast -}$ in nuclear matter is shown in
Fig.~\ref{fig:width-sig}.  The linear dependence of the decay widths
is caused by insensitivity of the elementary transition rate $\gamma
_{\SigStar N \to YN}$ to the Fermi motion of the initial nucleon.  For
the $\SigStar$-induced nonmesonic decays, there are several relations
due to the flavor SU(3) symmetry in the coupling constants.  In the
$\Sigma ^{\ast 0}$ case we obtain the same result for proton and
neutron in initial state because of the same coupling strengths in
the scattering amplitudes, hence we plot them in one figure 
[Fig.~\ref{fig:width-sig}(a)].  We also find that $\Gamma
_{\SigStarz N \to \Lambda N} / \Gamma _{\SigStarm p \to \Lambda
  n}=1/2$ and $\Gamma _{\SigStarz N \to \Sigma ^{\pm} N} / \Gamma
_{\SigStarm p \to \Sigma ^{0} n}=1$.

One interesting finding is that at all densities the
$\SigStar$-induced nonmesonic decay ratio $\Gamma _{\Lambda N}/\Gamma
_{\Sigma ^{0} N}$ is much larger than the $\LamStar$-induced one
$\approx 1.2$~\cite{Sekihara:2009yk}.  Especially in the
$\SigStarz$-induced case, we have very small branching ratio to the
$\Sigma ^{0}N$ final state.  This is caused by the small couplings
$\tilde{V}$ at both $\bar{K} N \Sigma ^{0}$ and $\eta NN$ vertices in
the $\Sigma ^{\ast 0}N\to \Sigma ^{0} N$ transition, hence the
$\SigStarz$ scarcely exchanges one Nambu-Goldstone boson for the
$\Sigma ^{0}$ final states.  Also it should be noted that there is no
relation between the $\Sigma ^{0}p$ ($\Sigma ^{0}n$) and $\Sigma
^{+}n$ ($\Sigma ^{-}p$) branching ratios, which should be $1/2$ if the
$I=0$ hyperon resonance appears in the initial state.  These points
will be important in the discussion of the $\LamStar / \SigStar$
contribution rate in the realistic kaon absorption experiments.

At the saturation density $\rho _{0}=0.17 \fm ^{-3}$, the total 
nonmesonic decay width is $43 \mev$ ($42 \mev$) for $\Sigma ^{\ast 0}$ 
($\Sigma ^{\ast -}$), in which 
$\Gamma _{\Lambda p}+\Gamma _{\Lambda n} = 25 \mev$, 
$\Gamma _{\Sigma ^{0} p}+\Gamma _{\Sigma ^{0} n} = 0.4 \mev$, and 
$\Gamma _{\Sigma ^{+} n}+\Gamma _{\Sigma ^{-} p} = 17 \mev$ 
($\Gamma _{\Lambda n} = 25 \mev$, 
$\Gamma _{\Sigma ^{0} n} = 9 \mev$, 
$\Gamma _{\Sigma ^{-} p} = 3 \mev$, and 
$\Gamma _{\Sigma ^{-} n} = 5 \mev$).  They are similar to 
the mesonic $\SigStar$ decay width in 
vacuum $\approx 37 \mev$.

\subsection{$\bm{\SigTEF}$ contribution to antikaon absorption}

Let us evaluate how the $\SigStar$ contributes the $K^{-}$ absorption
in nuclear matter.  For this purpose, we add coherently the $\SigStar$
contribution in the simple Breit-Wigner form as,
\begin{equation}
{\cal T}_{\pi Y}^{(\SigStar )} ( W ) = 
( \bm{p}_{\pi} \cdot \bm{S}_{1} ) 
\frac{D_{\pi Y} D_{K^{-} N_{1}}}{W - M_{\SigStar} + i \Gamma _{\SigStar}/2} 
( \bm{p}_{K^{-}} \cdot \bm{S}_{1}^{\dagger} ) , 
\end{equation}
for mesonic absorption and, 
\begin{align}
& {\cal A}^{(\SigStar )} ( K^{-} N_{1} N_{2} \xrightarrow{a} A B ) 
\nonumber \\ 
& = 
 \frac{\tilde{V}_{aN_{2}B} D_{a A} D_{K^{-} N_{1}}}
{W - M_{\SigStar} + i \Gamma _{\SigStar}/2} 
\tilde{\Pi} _{a}^{(p)} 
(q_{a}^{2}; \, \bm{q}, \, \bm{S}_{1}, \, \bm{\sigma}_{2}) 
( \bm{p}_{K^{-}} \cdot \bm{S}_{1}^{\dagger} ) , 
\end{align}
for nonmesonic absorption.  Here $M_{\SigStar}=1385 \mev$ and 
$\Gamma _{\SigStar}=37 \mev$ are mass and decay width of 
$\SigStar$, respectively, and the subscript $1$ in the 
$\bm{S}^{(\dagger )}$ denotes to be sandwiched by the spinors 
for $N_{1}$ and $A$.

\begin{figure*}[!Ht]
  \centering
  \begin{tabular*}{\textwidth}{@{\extracolsep{\fill}}cc}
    \Psfig{8.6cm}{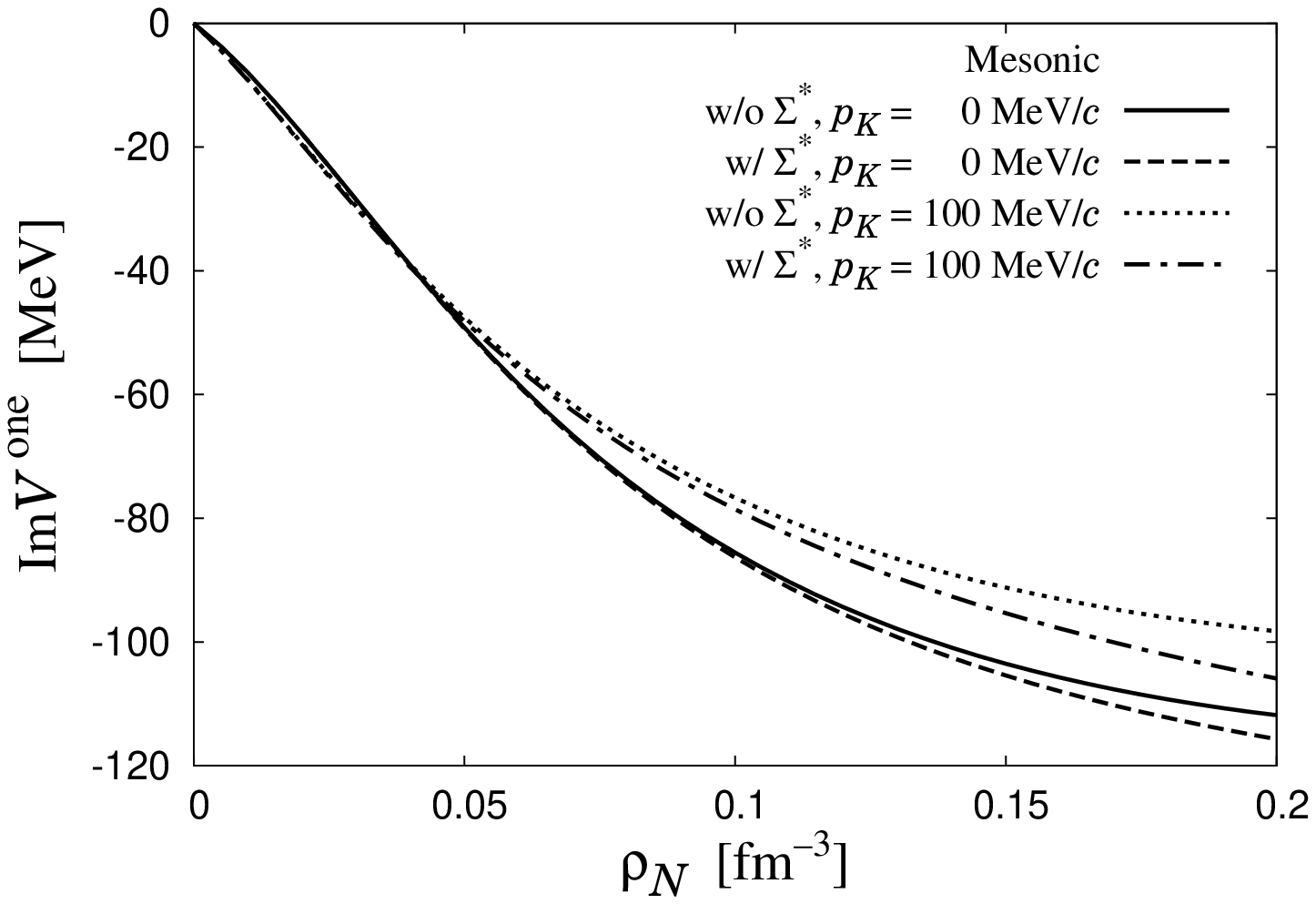} & 
    \Psfig{8.6cm}{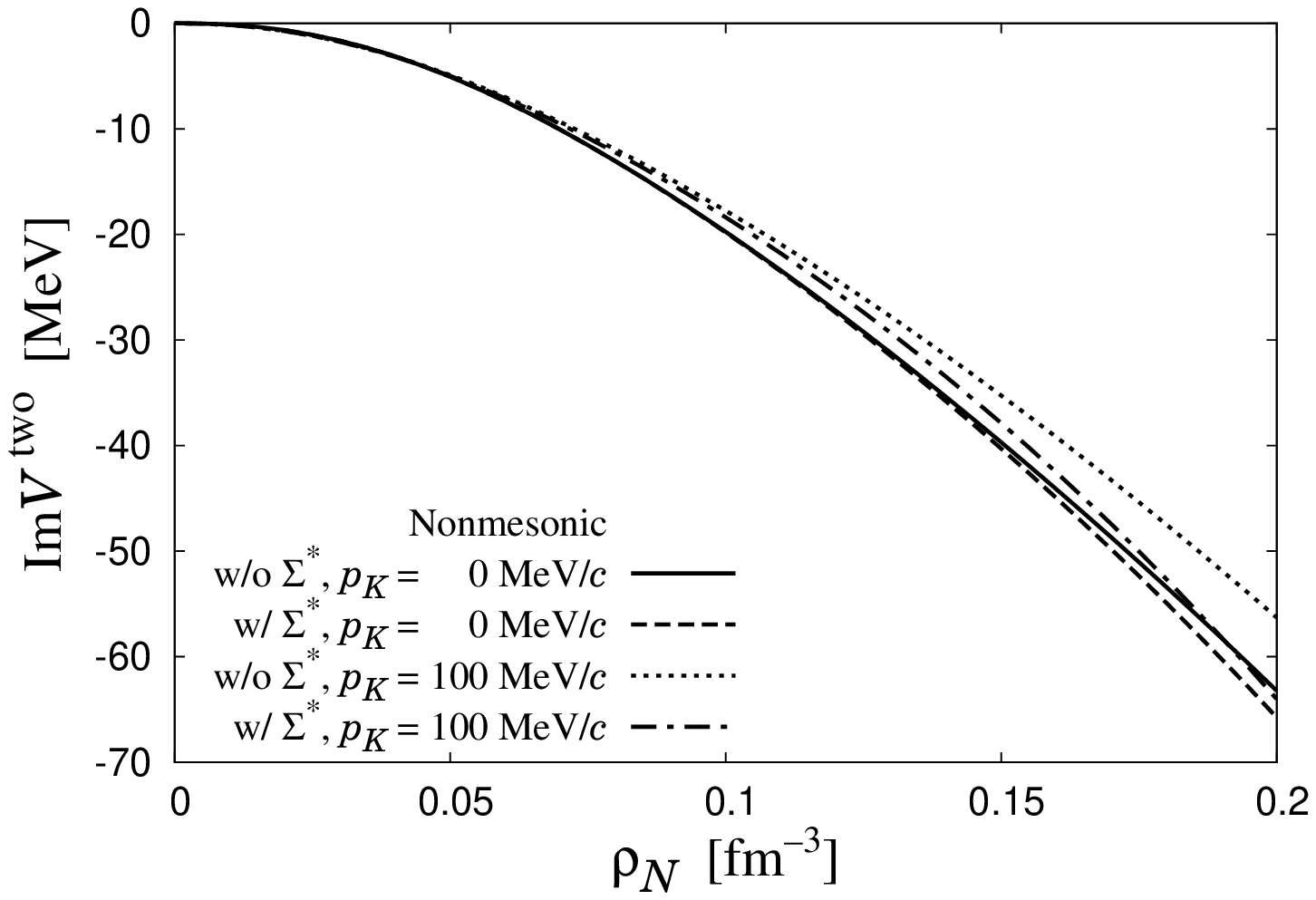} \\
    (a) & 
    (b)
  \end{tabular*}
  \caption{Total mesonic and nonmesonic absorption potentials
    [$\text{Im} V^{\text{one}}$ (a) and $\text{Im} V^{\text{two}}$
    (b), respectively] for $K^{-}$ in nuclear matter as functions of
    nuclear density.  Here we plot without and with $\SigStar$
    contributions in the $K^{-}N\to MB$ amplitude, and we take the
    kaon momentum $p_{K^{-}}=0\mev /c$ and $100 \mev /c$.  }
  \label{fig:V-SP}
\end{figure*}

The results for kaon energy-momentum $p_{K^{-}}^{\mu}=(m_{\bar{K}}, \,
\bm{p}_{K^{-}})$ with momenta $p_{K^{-}}\equiv \bm{p}_{K^{-}}=0 \mev
/c$ and $100 \mev /c$ are shown in Fig.~\ref{fig:V-SP}.  As we can
see, the $\SigStar$ contributions are constructively added to the
absorption potential.  However, the values of the contribution to the
potential are quite small compared with the $\LamStar$ one for the
kaon at rest, $p_{K^{-}}=0 \mev /c$, and even for $p_{K^{-}}=100 \mev
/c$ the shift of the absorption potential at the nuclear saturation
density is less than $10 \mev$ both in mesonic and nonmesonic cases.
This feature has been seen also in the $\bar{K}$-nucleus bound
systems~\cite{Weise:2008aj, YamagataSekihara:2008ji} and the analysis
of the kaonic atoms data~\cite{Cieply:2011fy}.  This is because
$\SigStar$ sits energy farther below the $\KbarN$ threshold than
$\LamStar$ and $\SigStar$ exists in $p$ wave of $K^{-}N$ system and
hence requires high momentum transfer, which is not adequately
achieved with slow $K^{-}$ and Fermi momentum of $N$.  Thus, we can
neglect the $\SigStar$ contribution to the absorption of slow $K^{-}$.

\section{Summary}
\label{sec:summary} 

In this paper we have theoretically investigated the branching ratios
of mesonic and nonmesonic $K^{-}$ absorption in nuclear matter in
order to understand the mechanism of $K^{-}$ absorption in experiments
by systematic evaluation of the decay patterns of $\bar{K}$-nucleus
systems from the low-energy $\KbarN$ interaction.  For the $K^{-}$
absorption, we have paid attention to two hyperon resonances,
$\LamStar$ and $\SigStar$, which are both below and close to the
$\KbarN$ threshold and thus will play important roles in the
absorption process.  The mesonic and nonmesonic absorption is
evaluated from the $K^{-}$ self-energy with one- and two-nucleon
interactions, respectively, which are the most probable contributions
at moderate nuclear densities.

As a result, within $s$-wave $\KbarN$ scatterings determined by the
chiral unitary approach, which dynamically generates $\LamStar$, we
have seen that both the mesonic and nonmesonic $K^{-}$ absorption
potentials at rest are dominated by the $\LamStar$ doorway process in
the $K^{-}p\to MB$ scattering.  The density dependence of the $K^{-}$
absorption potential shows non-$\rho^{1}_{N}$ (non-$\rho ^{2}_{N}$)
dependence due to the existence of the $\LamStar$ resonance in mesonic
(nonmesonic) absorption process.  We have found that the interference
between $\LamStar$ and the nonresonant $I=1$ background modifies
transition strengths of $K^{-}p$ to $\pi ^{+} \Sigma ^{-}$, $\pi ^{-}
\Sigma ^{+}$, and $\pi ^{0} \Sigma ^{0}$ channels below the threshold
and this modification can explain the ratios $[\pi ^{-} \Sigma ^{+}] /
[\pi ^{+} \Sigma ^{-}]$ ($R_{+-}$) of the branching ratios observed in
several kaonic atoms in experiments.  Due to the $\LamStar$ dominance
doorway process, the nonmesonic absorption ratios $[\Lambda p]/
[\Sigma ^{0} p]$ and $[\Lambda n]/ [\Sigma ^{0} n]$ are about unity
while $[\Sigma ^{+}n]/[\Sigma ^{0}p]$ and $[\Sigma ^{-}p]/[\Sigma
^{0}n]$ are about two.  Our approach gives that the mesonic and
nonmesonic absorption fractions are respectively about $70 \%$ and $30
\%$ at the saturation density, and with some $K^{-}$ atomic wave
functions and the Woods-Saxon density distribution we obtain the
fraction $\sim 10$--$20 \%$ for the nonmesonic absorption.  Taking
into account the kaon momenta and the energies, the absorption
potentials become weaker due to the downward shift of the initial
$K^{-}N$ two-body energy, but this does not drastirally change the
nonmesonic fraction.

We note that the density dependence of the decay pattern will be
realized by using nuclei with different atomic numbers as targets of
stopped $K^{-}$ reaction.  Especially the light nuclei such as the
deuteron, $\HeT$, and $\HeF$ will be suitable for this purpose, since
they serve as environment of various nuclear densities inside nuclei
due to the large varieties of the binding energies per one nucleon.

From the discussions on the $\SigStar$ contribution we have observed
different branching ratios and the larger total width in the
$\SigStar$-induced nonmesonic decay, where one $\SigStar$ exists in
nuclear medium in its initial state, compared with the
$\LamStar$-induced one discussed in the previous
study~\cite{Sekihara:2009yk}.  This fact will be important in the
discussion on $\LamStar / \SigStar$ contribution rate in the realistic
$\bar{K}$ absorption experiments.  In the slow $K^{-}$ absorption up
to the momentum $100 \mev/c$, however, $\SigStar$ has very small
contributions to the absorption process, because $\SigStar$ exists in
$p$ wave of the $K^{-}N$ system and requires high momentum transfer,
which is not adequately achieved with slow $K^{-}$ and the Fermi
momentum of $N$.  As a consequence, $\LamStar$ in the $s$-wave
$K^{-}p$ system gives dominant contributions to slow $K^{-}$
absorption.

\begin{acknowledgments}
We acknowledge 
A.~Ohnishi and 
K.~Yazaki 
for useful discussions. This work is partly supported by the 
Grand-in-Aid for Scientific Research from MEXT and JSPS 
(Nos. 
22740161, 
22-3389, 
and 24105706
), and 
the Grant-in-Aid for the Global COE Program ``The Next Generation of
Physics, Spun from Universality and Emergence'' from MEXT of Japan.
One of the authors, T.S. acknowledges the support by the Grand-in-Aid
from JSPS.  This work is
part of the Yukawa International Program for Quark-Hadron Sciences
(YIPQS).
\end{acknowledgments}

\end{document}